\documentclass[fleqn,usenatbib,useAMS]{mnras}
\usepackage{xfrac}
\usepackage{xcolor}
\usepackage{array}
\usepackage{amsmath,amssymb,amsfonts}
\usepackage{latexsym}
\usepackage{epstopdf} 
\usepackage{wasysym}
\usepackage{float}
\usepackage{array,multirow}
\usepackage{booktabs}
\usepackage{geometry}
\usepackage[strict]{changepage}
\usepackage{subfloat}
\usepackage{mathptmx}

\usepackage[T1]{fontenc}

\newcommand{\appropto}{\mathrel{\vcenter{
			\offinterlineskip\halign{\hfil$##$\cr
				\propto\cr\noalign{\kern2pt}\sim\cr\noalign{\kern-2pt}}}}}

\title[Evolving Binary Systems that Produce Novae]{In-depth Analysis of Evolving Binary Systems that Produce Nova Eruptions}

\author[Y. Hillman]{
Yael Hillman,$^{1}$\thanks{E-mail: yaelhi@ariel.ac.il}
\\
$^{1}$Department of Physics, Ariel University, Ariel, POB 3, 4070000, Israel\\
}
\date{Accepted XXX. Received YYY; in original form ZZZ}

\pubyear{2021}
\begin{document}
	\label{firstpage}
	\pagerange{\pageref{firstpage}--\pageref{lastpage}}
	\maketitle
		
\begin{abstract}
This study is the direct continuation of a previous work performed by Hillman et al., where they used their feedback dominated numerical simulations to model the evolution of four initial models with white dwarf (WD) masses of $0.7$ and $1.0M_\odot$ and red dwarf (RD) masses of $0.45$ and $0.7M_\odot$ from first Roche-lobe contact of the donor RD, over a few times $10^9$ years, until the RD was eroded down to below $0.1M_\odot$. This study presents an in-depth analysis of their four models complimented by three models with a higher WD mass of $1.25M_\odot$, one of which comprises an oxygen-neon (ONe) core. Common features were found for all seven models on a secular time scale as well as on a cyclic time scale. On the other hand, certain features were found that are strongly dependent either on the WD or the RD mass but are indifferent to the other of the two. Additionally, a model with a WD composed of an ONe core was compared with its corresponding carbon oxygen (CO) core WD model and found to have a significant impact on the heavy element abundances in the ejecta composition.
\end{abstract}


\section{Introduction}\label{sec:intro}
Novae are bright eruptions \cite[]{Pay1957} that occur in the degenerate envelope near the surface of a white dwarf (WD) as a result of accretion of hydrogen rich matter from a donor companion. The accumulating matter slowly raises the subsurface pressure and temperature until eventually igniting the accreted hydrogen in the CNO cycle \cite[]{Shara1981,MacDonald1983}. Since the degeneracy pressure is insensitive to temperature, heating does not cause the envelope to expand, thus, the temperature continues to rise, triggering a thermonuclear runaway (TNR) ejecting the accreted matter \cite[]{Starrfield1972}. It has been established that the manner in which a single nova eruption evolves may vary immensely depending on the system properties, specifically the WD mass and the rate that it accretes mass from its companion \cite[e.g.,][]{Shara1986,Kovetz1988,Livio1988,Kolb2001,Yaron2005,Toonen2014,Hillman2020}. In some cases, these properties may be deduced from observations \cite[e.g.,][]{Schaefer2010,Strope2010,Shara2018}. A nova producing system undergoes extensive secular changes as well. \cite{Hillman2020} have shown that the donor may be eroded over a few Gyr as the result of thousands of nova cycles of accretion and eruption, while the mass of the WD changes very little. They demonstrated this by using a feedback-dominated code designed to follow the evolution of the stellar binary components from an initial state of Roche lobe overflow (RLOF) of a red dwarf (RD) donor onto the surface of a WD accretor. They showed that the mass transfer rate ($\dot{M}$) from the RD to the WD, which had been assumed constant in nearly all earlier simulations \cite[]{Prikov1995,Idan2013,Wolf2013,Hillman2015}, varies over a few orders of magnitude --- rapidly decreasing and then slowly increasing --- during the accretion phase of a single nova cycle i.e., between eruptions. They have also found that $\dot{M}$ evolves slowly over many orders of magnitude throughout the course of thousands of nova cycles, ranging from as high as $\sim10^{-7}M_\odot yr^{-1}$ to low enough as to be considered effectively zero. Different rates of accretion characterize different types of eruptive behavior \cite[]{Collazzi2009,Knigge2011,Dubus2018}, i.e., classical novae (CN), recurrent nova (RN), dwarf nova (DN) and nova-likes (NL). 
\cite{Hillman2020} showed that the behavior of their four different models, spanning a vast range of accretion rates throughout evolution, implies that the various types of phenomena do not originate from different types of systems, but rather occur at different evolutionary phases of the same type of system, demonstrating that their models alternate between states of being detached hibernating systems, NLs, DNs and CNs depending on the accretion rate at the time. This paper is a direct continuation of their study, 
presenting an in-depth analysis of their results. In addition, included here are three more models as an effort to expand this exercise to consider more massive WDs, and a WD with an oxygen-neon (ONe) core. This study investigates the trends common to all the models, as well as the differences between the models, focusing on determining the key parameters that dominate each of these trends. This includes short term features of the nova cycle that characterize the eruption phase as well the accretion phase, and secular changes over thousands of cycles that slowly erode the RD.  \S\ref{sec:models} describes the seven models and the methods of calculation. The results are in \S\ref{sec:results} followed by a comparison with the results of an ONe core model in \S\ref{sec:ONe}. The conclusions are summarized in \S\ref{sec:conclusions}

\section{Methods and models}\label{sec:models}
All the simulations were carried out with the combined code for simulating the long-term evolution of a binary system producing consecutive nova eruptions \cite[]{Hillman2020}. This code utilizes two originally independent codes that have been modified and combined. 
The first simulates the donor component by the use of a hydrostatic stellar evolution code which is designed to follow the evolution of a star from pre-MS all the way through to a WD \cite[]{Kovetz2009}. The data produced by this code was used to build a dense database of donor input parameters to be used in the second code --- a hydrodynamic Lagrangian nova evolution code designed to follow the evolution of hundreds of thousands of complete consecutive nova cycles on the surface of a WD \cite[]{Prikov1995,Epelstain2007,Hillman2015}. \cite{Kalomeni2016} found that regardless of the initial RD age, the ratio of orbital period ($P\rm_{orb}$) to RD mass ($M_{RD}$) for cataclysmic variables (CV) in RLOF remains the same. Since a CV system can emerge in the state of RLOF after numerous different scenarios, for simplicity, in this work each simulation begins with a zero age main sequence (ZAMS) donor filling its Roche lobe (RL). The combined code calculates the rate of mass accretion onto the surface of the WD at each timestep by accounting for orbital momentum change due to magnetic braking (MB) and gravitational radiation (GR) following \cite[]{Paxton2015}:

\begin{equation}\label{eq:JMB}
\dot{J}_{MB}=-1.06\times 10^{20}M_{RD}R_{RD}^4P\rm_{orb}^{-3}
\end{equation}  
\begin{equation}\label{eq:JGR}
\dot{J}_{GR}=-\frac{32}{5c^5}\left(\frac{2\pi G}{P\rm_{orb}}\right)^\frac{7}{3}\frac{(M_{RD}M_{WD})^2}{(M_{RD}+M_{WD})^\frac{2}{3}}
\end{equation}
and by accounting for the separation change due to mass lost from the system during each nova eruption by using:
\begin{equation}\label{eq:delta_a}
\Delta{a}=2a\left(\frac{m{\rm_{ej}}-m{\rm_{acc}}}{M_{WD}}+\frac{m\rm_{acc}}{M_{RD}}\right)
\end{equation}
where $\dot{J}_{MB}$ and  $\dot{J}_{GR}$ are the change in orbital angular momentum due to magnetic braking (MB) and gravitational radiation (GR) respectively, $M_{WD}$ and $M_{RD}$ are the masses of the WD accretor and the RD donor respectively, $R_{RD}$ is the radius of the RD, $P\rm_{orb}$ is the orbital period of the binary system, and $m\rm_{acc}$ and $m\rm_{ej}$ are the accreted and ejected masses of the previous nova cycle respectively.
Further description of the method, including the treatment of the irradiated RD for tens to hundreds of years after each nova eruption, may be found in \cite{Hillman2020}, noting that all processes are treated as spherically symmetrical. The implications of this simplification on the irradiated RD means that the increase in the RD's effective temperature on the side near the WD may be higher than calculated here \cite[]{Kovetz1988}, which may increase the enhanced RLOF and therefore, the mass transfer rate in the tens to hundreds of years following an eruption. Justification for treating the effects of irradiation as spherical is based on the calculations of \cite{Kovetz1988} who showed that the thermal timescale is of order $\sim10^4$ seconds, which is very much shorter than the timescale of tens to hundreds of years during which the RD is irradiated due a nova eruption. The RD bloating is adopted from \cite{Kovetz1988} and then parameterized \cite[see][]{Hillman2020}  in order to accommodate for different WD masses.
A total of seven models were analyzed here. The four models ($\#$1 through $\#$4) that are presented in \cite{Hillman2020} in addition to three models ($\#$5 through $\#$7) with a more massive WD ($M_{WD}=1.25M_\odot$) one of which consists of an oxygen-neon (ONe) core. The initial model parameters are summarized in Table \ref{tab:mdls}.

\begin{table}
	\begin{center}
		\begin{tabular}{ccccc}
			\hline
			{$\#$}&{Model}&{WD type} &{$M\rm_{WD}[M_\odot]$}& {$M\rm_{RD}[M_\odot]$}\\
			\hline
			{1.}&{$070-070$}&{CO}& {0.70}& {0.70}\\
			{2.}&{$070-045$}&{CO}& {0.70}& {0.45}\\
			{3.}&{$100-070$}&{CO}& {1.0}& {0.70}\\
			{4.}&{$100-045$}&{CO}& {1.0}& {0.45}\\
			{5.}&{$125-070$}&{CO}& {1.25}& {0.70}\\
			{6.}&{$125-045$}&{CO}& {1.25}& {0.45}\\
			{7.}&{$125-045$ (ONe)}&{ONe}& {1.25}& {0.45}\\
			\hline					
		\end{tabular}
		\caption{Initial model parameters.\label{tab:mdls}}
	\end{center}
\end{table}

\section{Results}\label{sec:results} 

\begin{figure}
	\begin{center}
		{\includegraphics[trim={0.0cm 1.0cm 0.5cm 1.2cm},clip ,
			width=1.0\columnwidth]{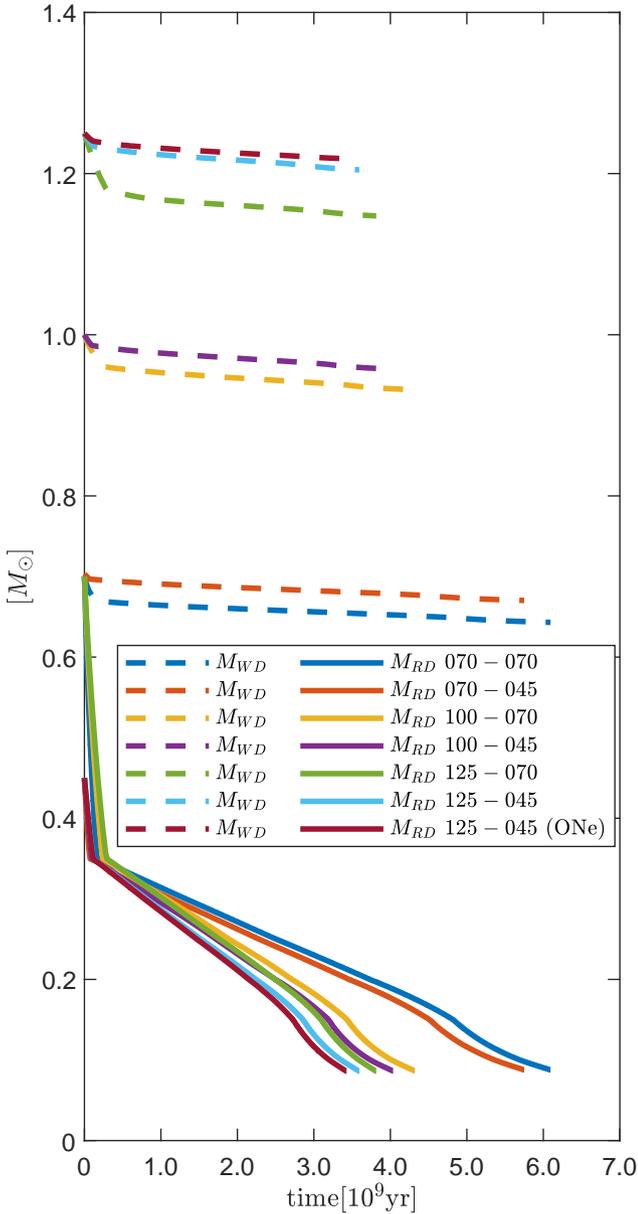}}
		\caption{\label{fig:Masses}Evolution of the WD and RD masses for the seven models.}
	\end{center}
\end{figure}

The evolution of the binary masses for the seven models are presented in Figure \ref{fig:Masses}, showing that the three new models with a $1.25M_\odot$ WD (models $\#$5 through $\#$7) behave in a manner similar to the previous four models with less massive WDs --- all of the WD masses remain almost constant while the WDs of models with a $0.45M_\odot$ RD donor lose less mass than the WDs (of equal initial masses) of models with a $0.7M_\odot$ donor. This is because during each nova eruption the WD ejects more mass than it has accreted during the previous accretion phase meaning that the net mass change at the end of each cycle is negative. Since $m\rm_{acc}$ is inversely proportional to $M_{WD}$ \cite[e.g.,][]{Yaron2005}, WDs with more massive companions require a larger number of cycles (accretion phases) in order to erode their RD companion, meaning that they experience more eruptions as well, resulting in a larger total net mass loss. 

\begin{figure*}
	\begin{center}
	{\includegraphics[trim={0.7cm 0.4cm 1.0cm 1.0cm},clip ,
		width=1.0\columnwidth]{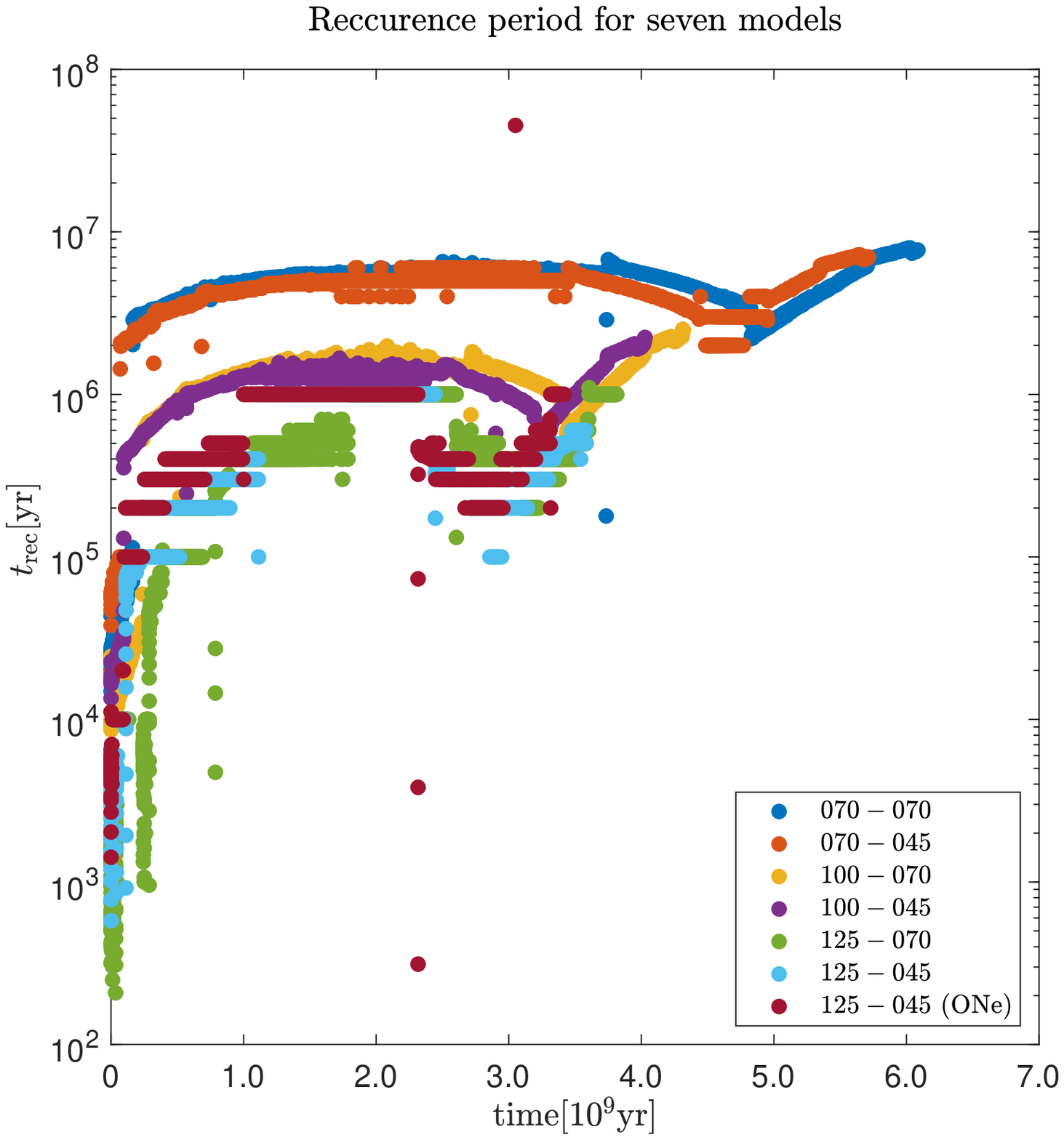}}
		{\includegraphics[trim={1.0cm 0.4cm 0.8cm 1.0cm},clip ,
			width=1.0\columnwidth]{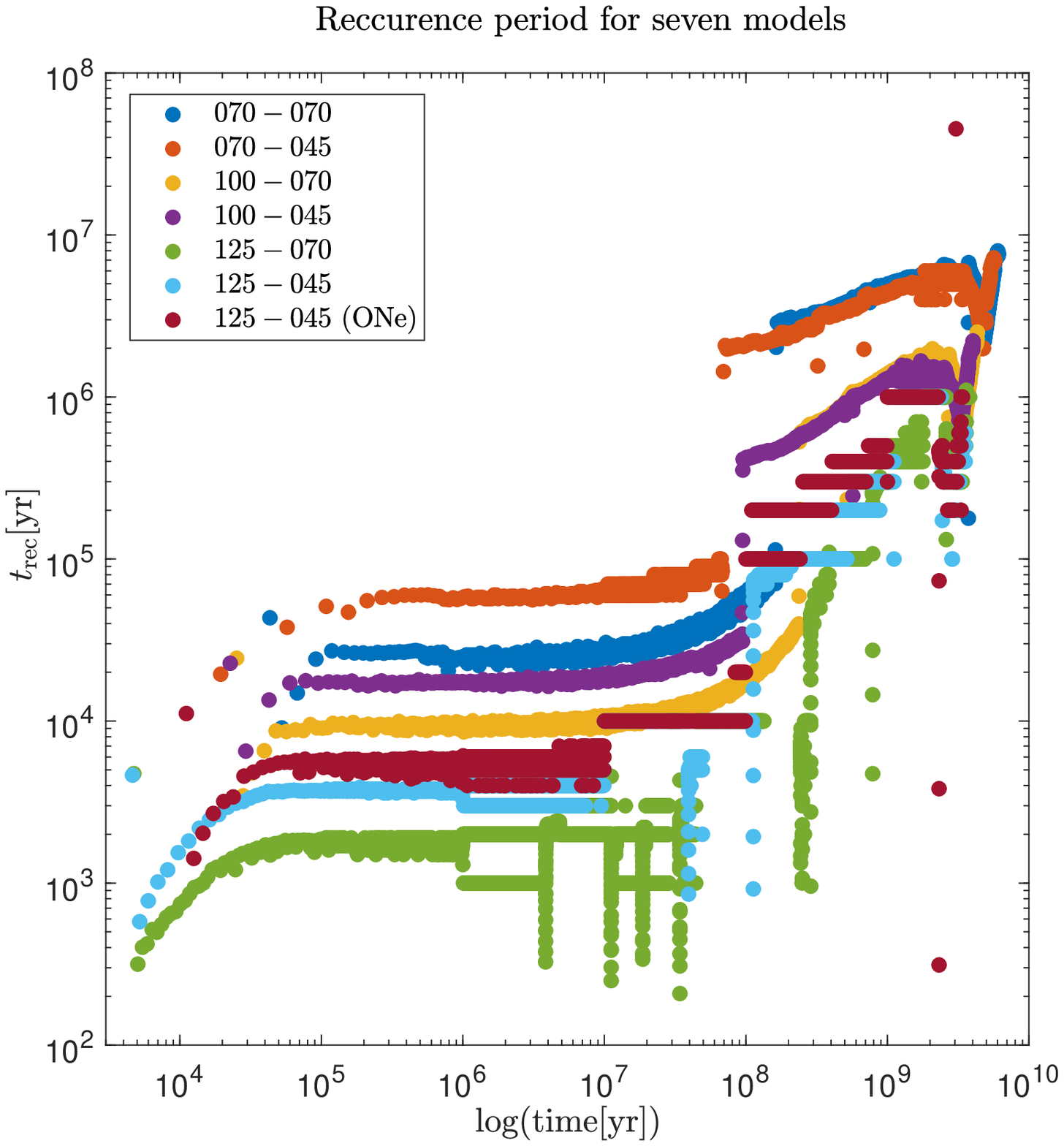}}
		\caption{\label{fig:trec}Recurrence time for the seven models, linear timescale (left) and logarithmic timescale (right).}
	\end{center}
\end{figure*}

\begin{figure*}
	\begin{center}
		{\includegraphics[trim={0.5cm 0.0cm 0.5cm 1.5cm},clip ,
			width=1.0\columnwidth]{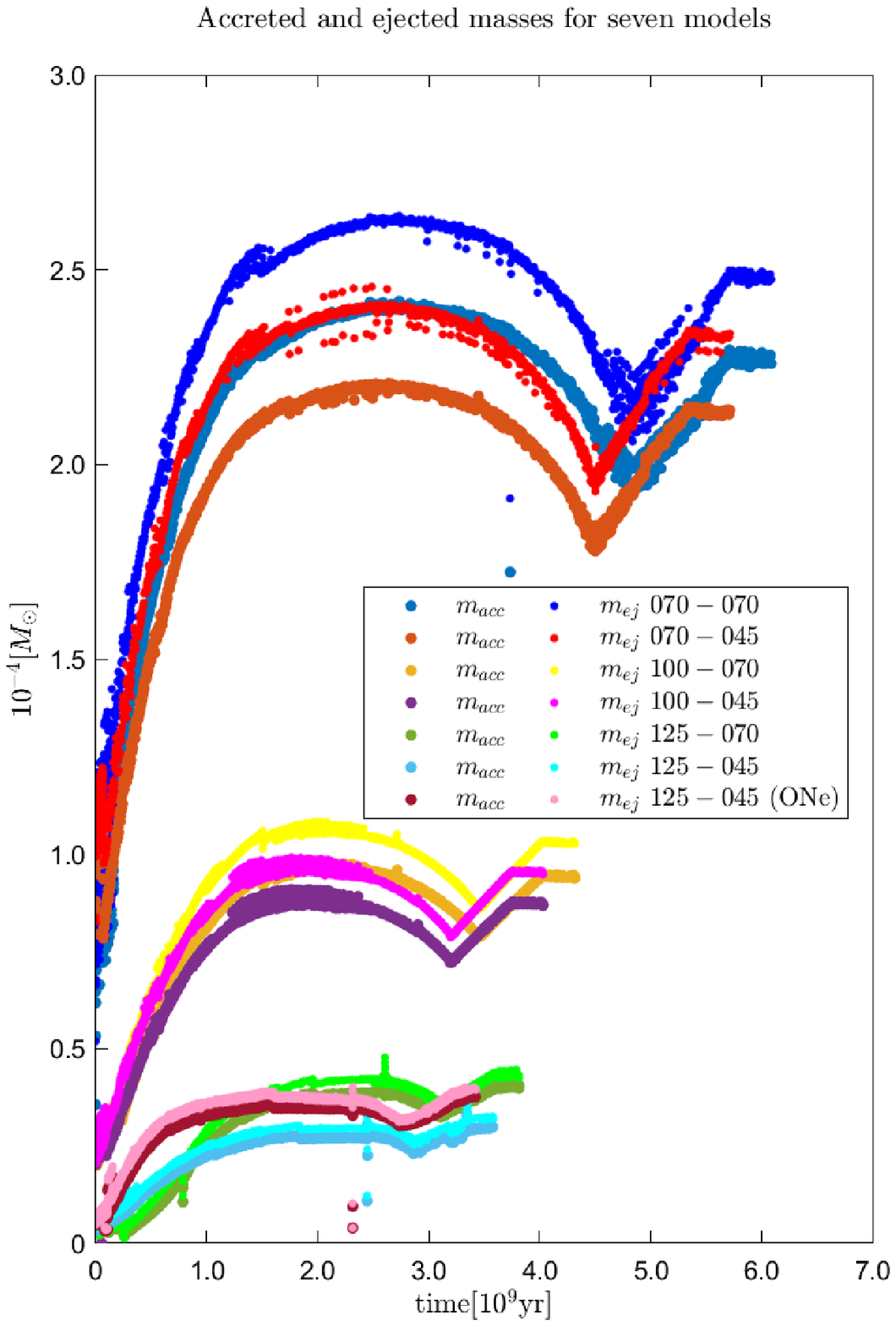}}
		%
		{\includegraphics[trim={0.5cm 0.0cm 0.5cm 1.5cm},clip ,
			width=1.0\columnwidth]{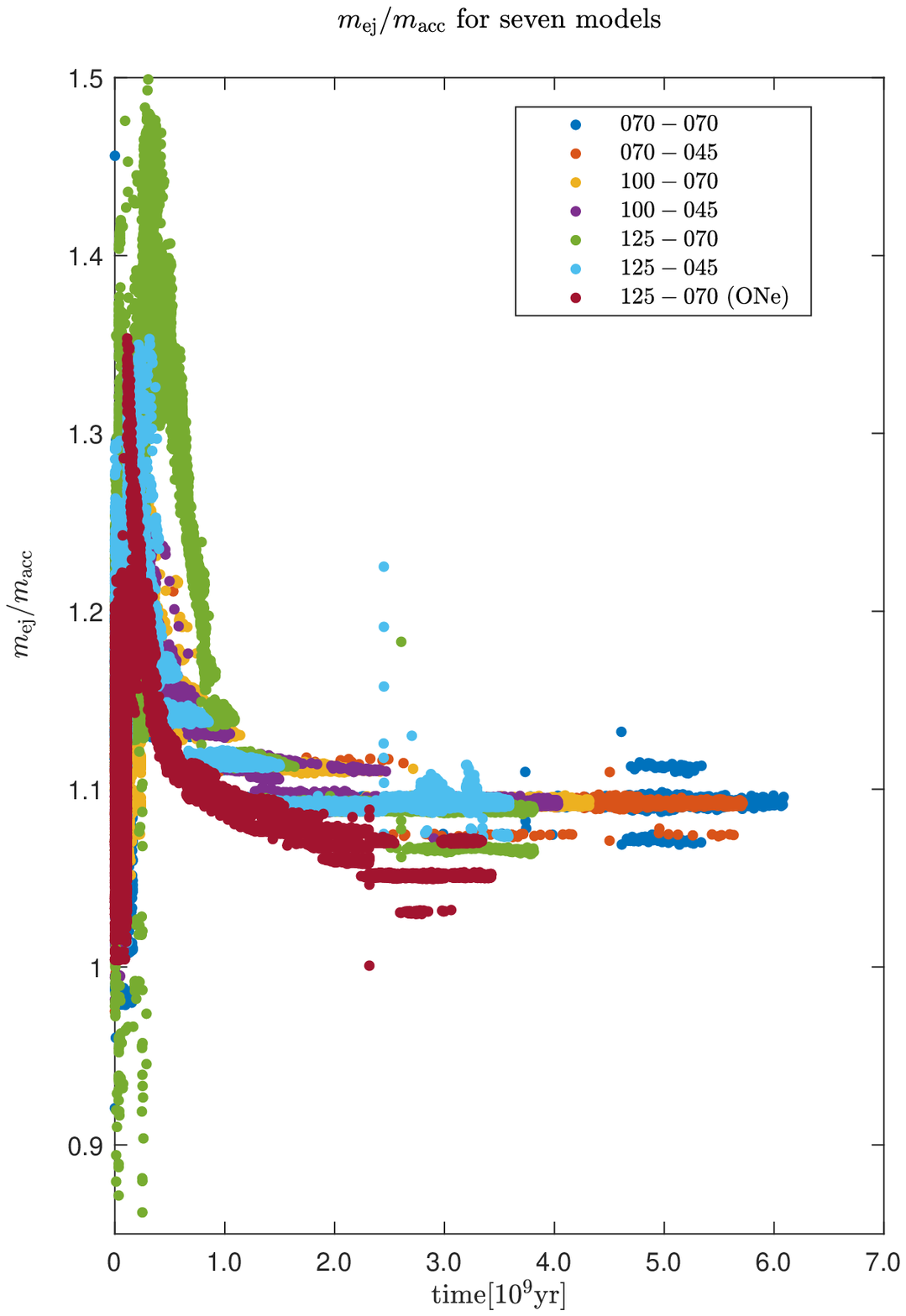}}
		\caption{Accreted and ejected masses (left) and the ratio $m{\rm_{ej}}/m{\rm_{acc}}$ (right) for the seven models.\label{fig:mej_macc}}
	\end{center}
\end{figure*}

Although the models with the more massive ($0.7M_\odot$) RDs  experience roughly a factor of two more cycles than do the models with the less massive ($0.45M_\odot$) RDs (when comparing between models with the same initial WD masses) the total evolutionary time is almost the same. This is because the time to erode a RD down to $\sim0.35M_\odot$ constitutes only a small fraction of the time it takes to continue eroding the RD down to $<0.1M_\odot$. Essentially, the total evolutionary time of any $M_{WD}+M_{RD}$ combination model may be determined by the time it takes for the WD to erode a $0.35M_\odot$ RD. This is the source of the "knee" in all the mass curves (WDs and RDs) in Figure \ref{fig:Masses}. The reason for this discrepancy before and after $M_{RD}\simeq0.35M_\odot$ lies in the effect of magnetic braking (MB). When the RD mass falls below $\sim0.35M_\odot$, for which RDs become fully convective, the MB becomes inefficient in removing angular momentum and the binary separation ($a$) decreases at a substantially slower rate. This causes the accretion rate to increase at a slower rate as well, considerably extending the time needed for the WD to accumulate the critical mass ($m\rm_{acc}$) that is needed for triggering a TNR that will lead to a nova eruption. It is noted that for high mass loss rates, roughly higher than a few times $10^{-10}M_\odot yr^{-1}$, the RD will become fully convective at a mass lower than 0.35$M_\odot$ (this is in agreement with e.g., \cite{Howell2001}). Since for RDs approaching 0.35$M_\odot$, the accretion rate does not become this high for most of the accretion time, this effect may only slightly decrease the total time of the accretion phase for these RDs until the RD is somewhat more eroded.

An additional aspect which may be seen in Figure \ref{fig:Masses} is the relation between the total evolutionary times of models with different WD masses. More massive WDs erode their donor down to $<0.1M_\odot$ faster. 
The total evolutionary time ($t$) may be estimated as the (average) recurrence time ($t\rm_{rec}$) multiplied by the number of cycles.
$t\rm_{rec}$ is given in Figure \ref{fig:trec}, showing a ten fold decrease in the $t\rm_{rec}$ of the  $1.25M_\odot$ WD models as compared with the $0.7M_\odot$ WD models, while the $t\rm_{rec}$ of $1.0M_\odot$ WD models are in between. The reason for this is because a more massive WD builds up the required critical sub-surface pressure with less accreted mass ($m{\rm_{acc}}$), thus, (for a given accretion rate) less time is needed. This also means that (for a given WD mass) the critical accreted mass required to trigger a TNR is lower as well, as may be seen in the left panel of Figure \ref{fig:mej_macc}, where the curves representing more massive WDs are lower than the curves representing less massive WDs. The ejected mass ($m\rm_{ej}$) shows a similar trend. On the other hand, the general shape of the curves is common to all seven models, expressing that the trend of $m\rm_{acc}$ and $m\rm_{ej}$ changes twice throughout evolution. They begin increasing, reaching a peak doubling their original value when the RD mass reaches roughly ${\sim0.3M_\odot}$. At this point there is a plateau after which they begin to decline. When the RD mass reaches ${\sim0.15M_\odot}$ they resume increasing. This is in correlation with the general trend of the accretion rate, and of $t\rm_{rec}$. The number of cycles may be estimated as $0.35M_\odot$ divided by the (average) $m\rm_{acc}$, giving the total evolutionary time as $t\propto {t\rm_{rec}}/{m\rm_{acc}}$. Although, $t\rm_{rec}$ and $m\rm_{acc}$ both decrease with increasing WD mass, so this simple parameterized calculation does not contribute sufficient information to the trend of the total evolutionary time. However,  ${t\rm_{rec}}/{m\rm_{acc}}\propto\dot{M}\rm_{avg}^{-1}$, means that, $t\propto\dot{M}\rm_{avg}^{-1}$ leading to the conclusion that a shorter total evolutionary time implies a higher total evolutionary average accretion rate. This means that on an evolutionary time scale, models with more massive WDs accrete at higher average rates and  will erode their RD companion faster, \textit{regardless of the RDs initial mass}. 

\begin{figure}
	\begin{center}
		{\includegraphics[trim={0.0cm 0.5cm 0.0cm 0.9cm},clip ,
			width=1.0\columnwidth]{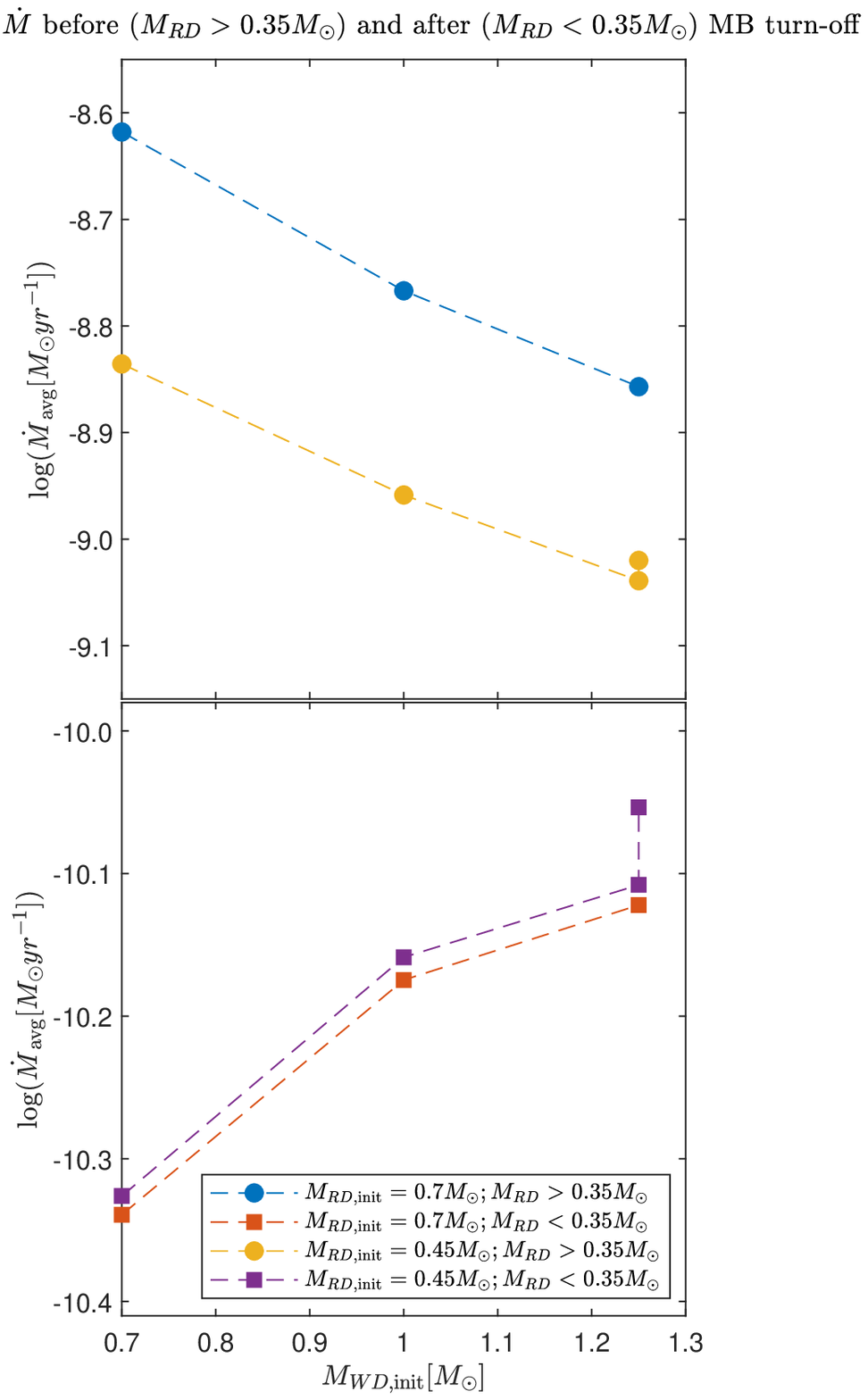}}
		\caption{The average accretion rate for each of the seven models before ($M_{RD}>0.35M_\odot$) and after ($M_{RD}<0.35M_\odot$) magnetic braking turn-off. \label{fig:Mdot_avg_bfr_aftr}}
	\end{center}
\end{figure}

The average accretion rate per model ($\dot{M}\rm_{avg}$), each divided into two epochs, is demonstrated in Figure \ref{fig:Mdot_avg_bfr_aftr}. The squares show $\dot{M}\rm_{avg}$ for the epoch of eroding the RD mass from $0.35M_\odot$ down to $<0.1M_\odot$, showing that a higher initial $M_{WD}$ yields a higher total $\dot{M}\rm_{avg}$ for this part of the evolution, for all the models. On the contrary, the circles representing $\dot{M}\rm_{avg}$  for the earlier evolutionary phase of eroding the donor from its initial mass down to $0.35M_\odot$, show an opposite trend --- the more massive initial WD masses have a slower $\dot{M}\rm_{avg}$. However this has a negligible influence on the total evolutionary time because the total time of this stage, for all seven models, is only a small fraction of the total evolutionary time.

\begin{figure*}
	\begin{center}
		{\includegraphics[trim={0.5cm 0.4cm 0.5cm 1.1cm},clip ,
			width=1.0\columnwidth]{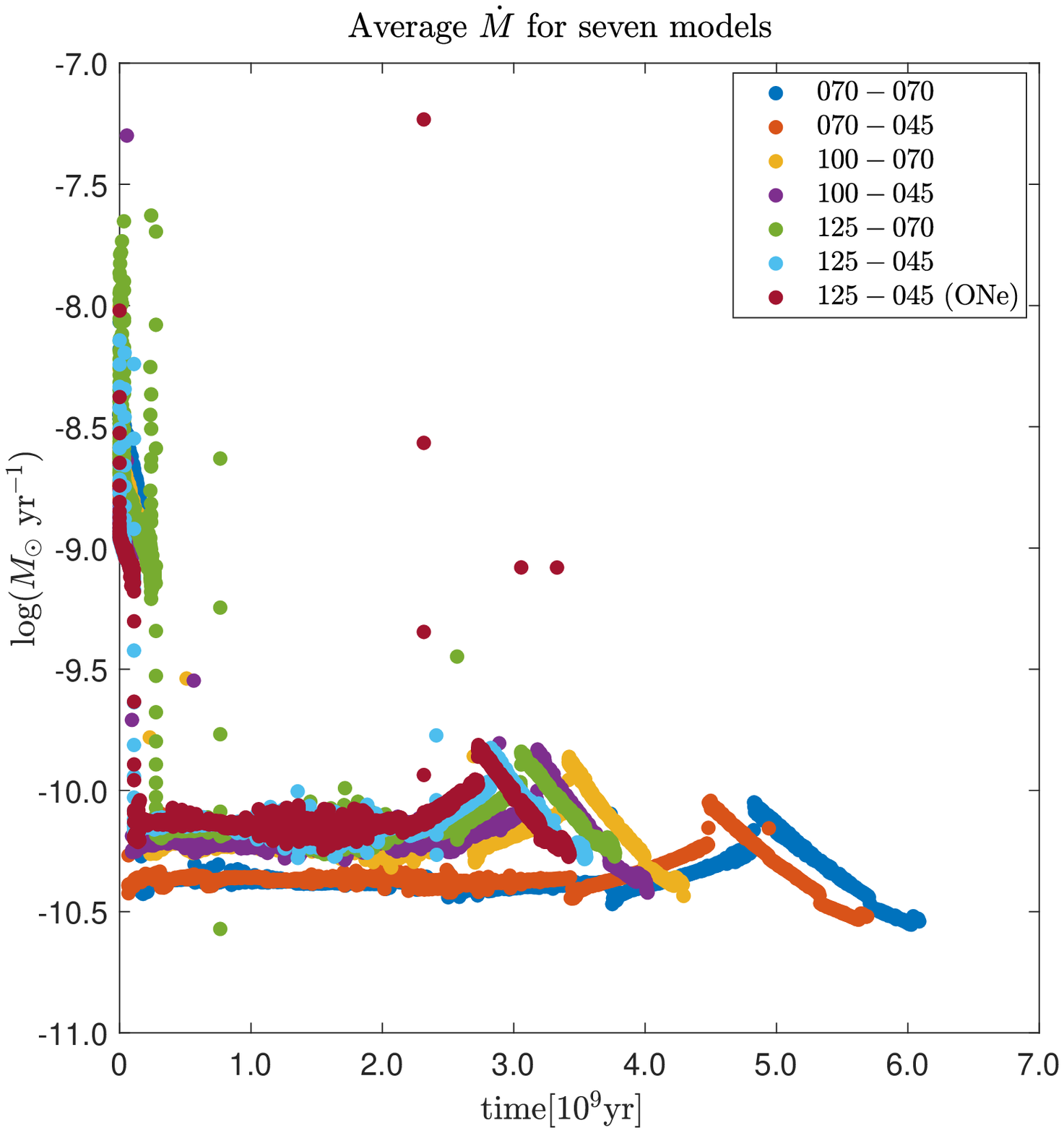}}
{\includegraphics[trim={0.5cm 0.4cm 0.5cm 1.1cm},clip ,
	width=1.0\columnwidth]{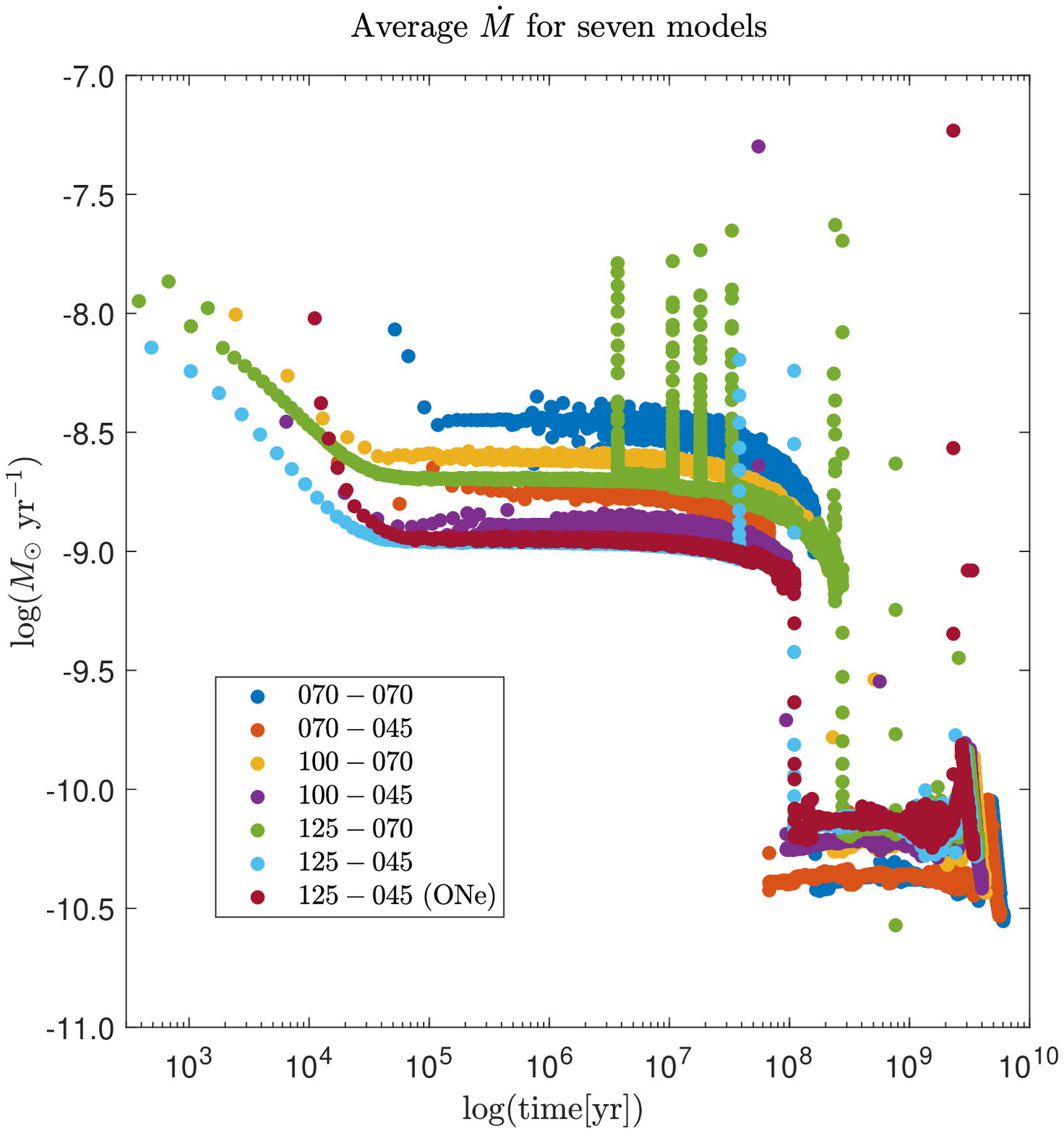}}
		\caption{The average accretion rate per cycle for the seven models, linear timescale (left) and logarithmic timescale (right).\label{fig:Mdot_avg}}
	\end{center}
\end{figure*}

A more detailed examination of the accretion rate, averaged per cycle is given in Figure \ref{fig:Mdot_avg}. The average accretion rate drops about two orders of magnitude when the MB stops, as expected from Figure \ref{fig:Mdot_avg_bfr_aftr}, but it also shows a more subtle change as the RD mass approaches ${\sim0.15M_\odot}$. At such low RD masses, hydrogen burning becomes less efficient, resulting in the radius not contracting with mass loss as efficiently as before. This causes the RLOF to increase, and thus causes $\dot{M}$ to temporarily stray away from its fairly constant secular average rate. It then decreases back to its normal value, until the RD is eroded ($M_{RD}<0.1M_\odot$). The period minimum (discussed later) occurs at $M_{RD}\simeq0.15M_\odot$ as well, for the same reason as the temporary $\dot{M}$ increase. 

\begin{figure}
	\begin{center}
	{\includegraphics[trim={0.0cm 1.2cm 0.0cm 0.7cm},clip , 			width=1.0\columnwidth]{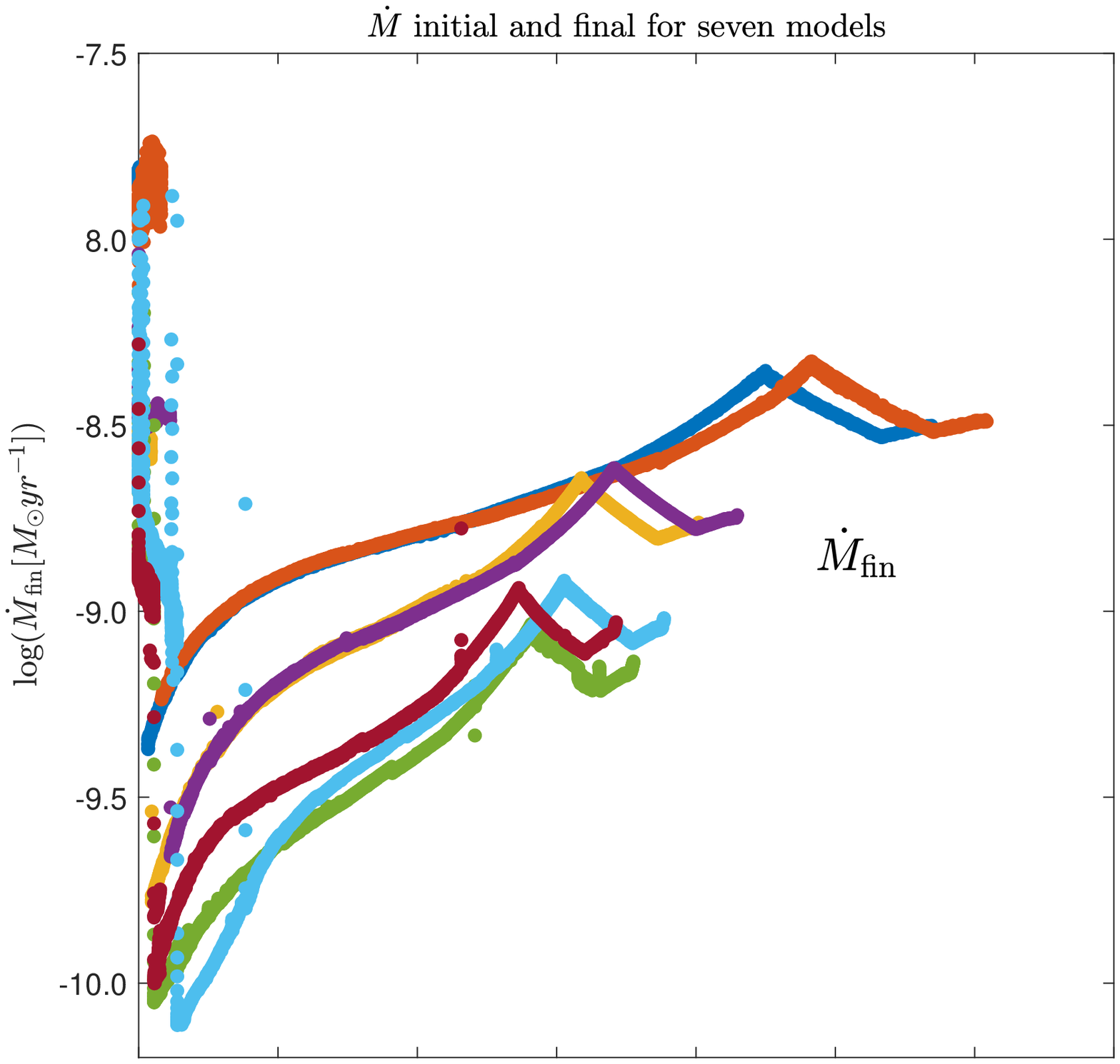}}
	{\includegraphics[trim={0.0cm 0.15cm 0.0cm 0.9cm},clip , 			width=1.0\columnwidth]{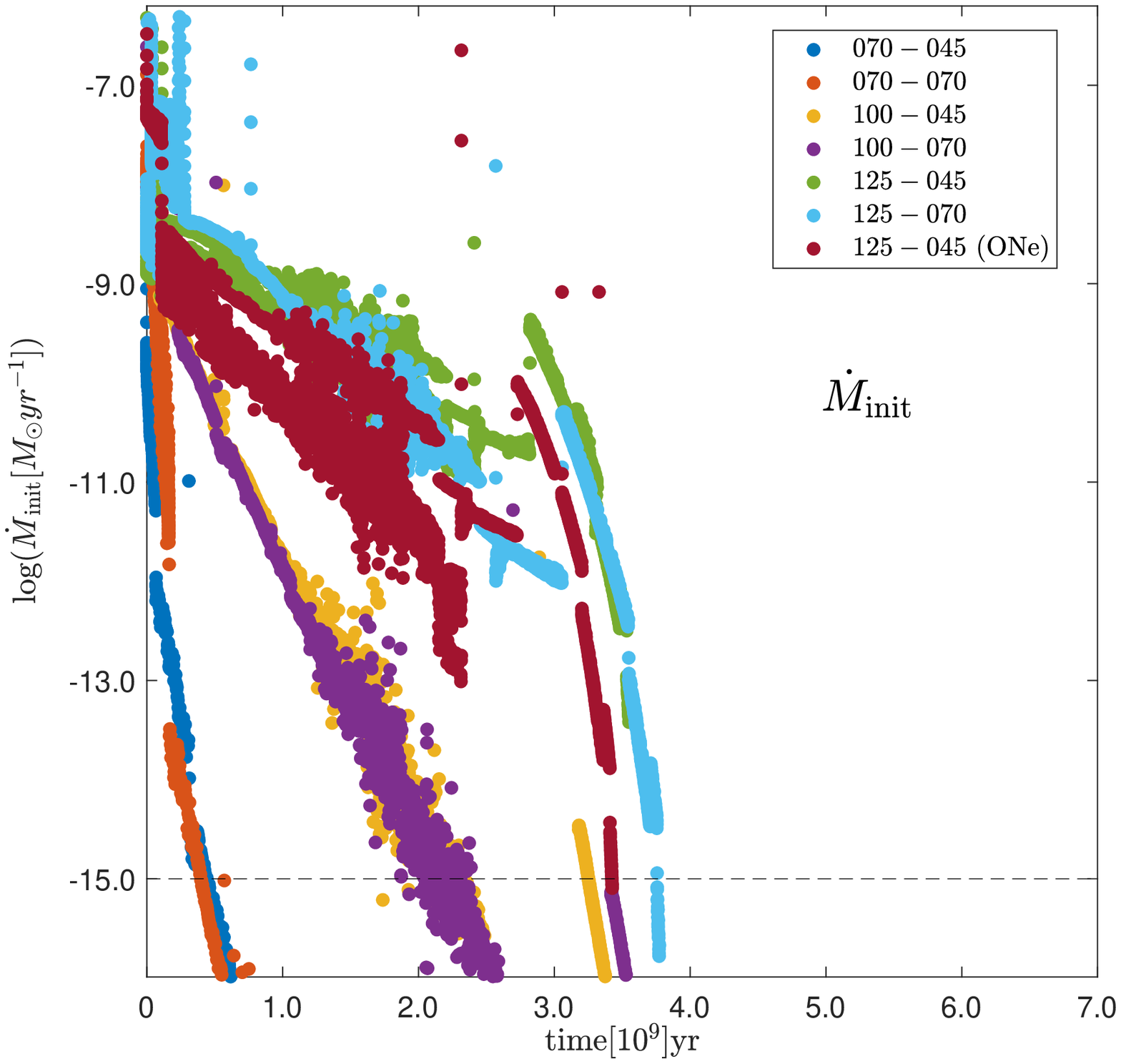}}
		\caption{The final (top) and initial (bottom) accretion rate per cycle for the seven models.\label{fig:Mdot_i_f}}
	\end{center}
\end{figure}

Remarkably, even though the average accretion rate per cycle (Figure \ref{fig:Mdot_avg}) \textit{and} per epoch (Figure \ref{fig:Mdot_avg_bfr_aftr}) are very similar for all the models, looking closely at the \textit{initial} and \textit{final} accretion rates shows diversity depending strongly on the WD mass. This is presented in Figure \ref{fig:Mdot_i_f} showing that for all the models the initial accretion rate per cycle ($\dot{M}\rm_{init}$) decreases and the final accretion rate per cycle ($\dot{M}\rm_{fin}$) increases seculary, however for the less massive WDs ($0.70M_\odot$) the changes occur faster resulting in these systems reaching a state of accretion rate which is practically zero ($\lesssim10^{-15}\dot{M}yr^{-1}$) at an earlier evolutionary point than the moderate mass WDs ($1.0M_\odot$), and the 
massive WDs ($1.25M_\odot$) erode their donor before $\dot{M}$ has a chance to become very low. The trend is the opposite for $\dot{M}\rm_{fin}$ because the longer the accretion phase, the longer time there is for orbital momentum loss to decrease the separation and thus increase $\dot{M}$, resulting in a higher $\dot{M}\rm_{fin}$ for a lower WD mass. Because the trends are opposite, the average rates (Figure \ref{fig:Mdot_avg}) are similar.


\begin{figure*}
	\begin{center}
		{\includegraphics[trim={0.5cm 0.0cm 0.5cm 1.1cm},clip ,
			width=1.0\columnwidth]{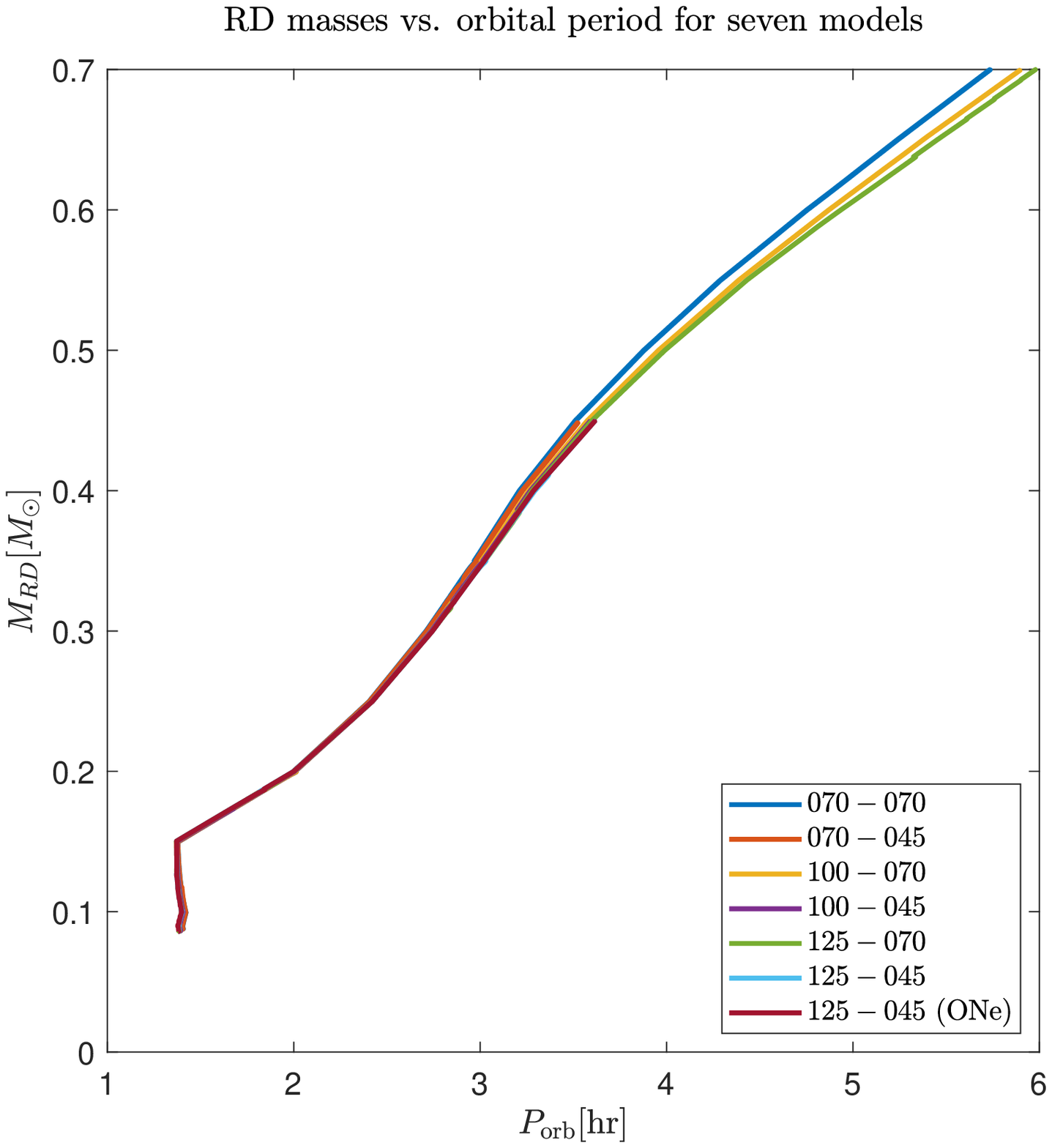}}
		{\includegraphics[trim={0.5cm 0.0cm 0.5cm 1.1cm},clip ,
			width=1.0\columnwidth]{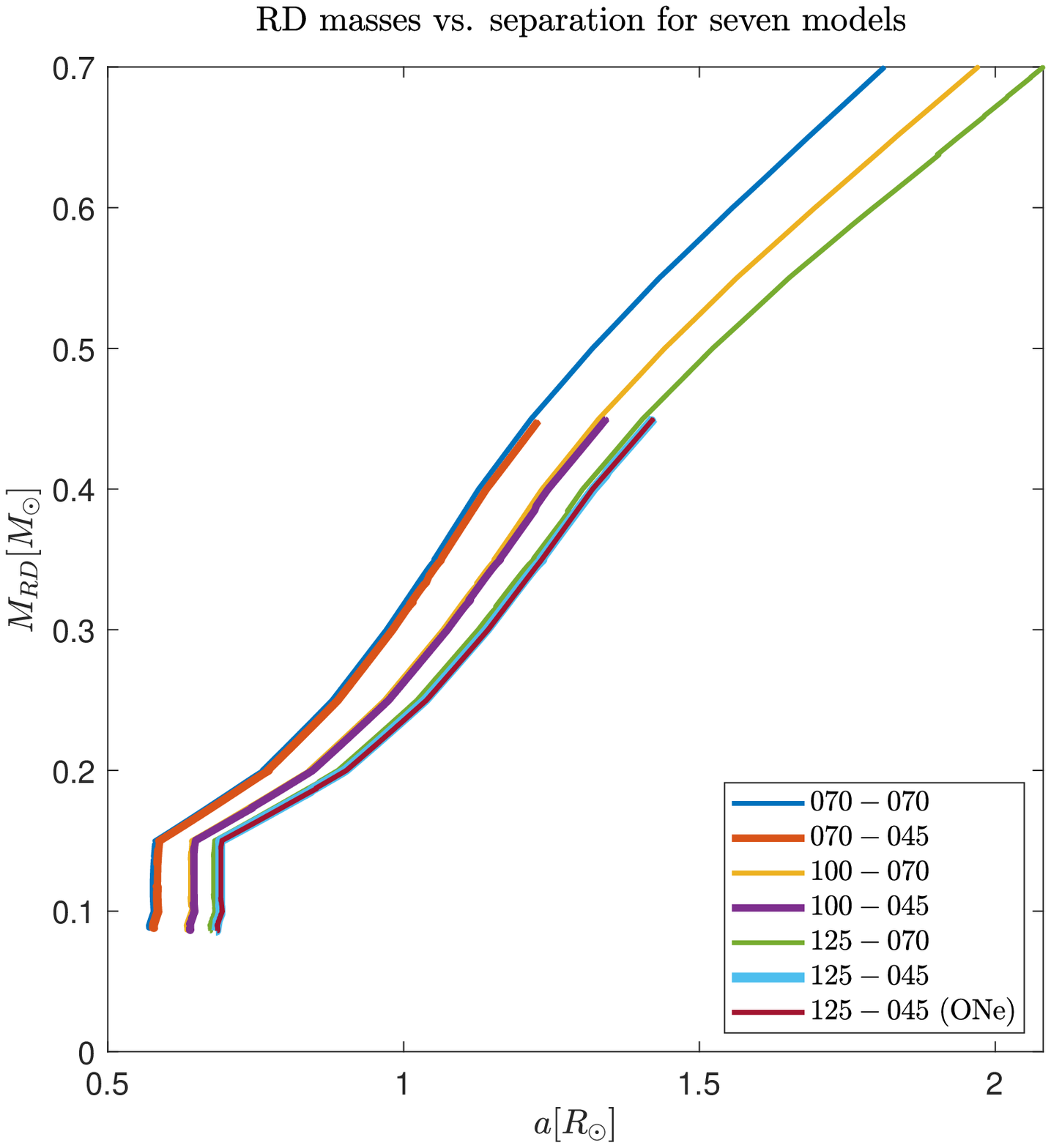}}%
		\caption{\label{fig:MRD_a_Porb}Donor mass evolution vs. orbital period (left) and binary separation (right) for the seven models.}
	\end{center}
\end{figure*}

The accretion rate is strongly dependent on the binary separation (and orbital period) \cite[]{Ritter1988,Livio1994b,Kolb2001,Hillman2020} which show a clear trend in Figure \ref{fig:MRD_a_Porb} --- the RD mass determines $P\rm_{orb}$. All three of the models with a $0.7M_\odot$ RD begin the simulation with an orbital period of $\sim6$ hours, regardless of the WD mass (the curves overlap) and the period decreases with decreasing RD mass. When these RD masses are eroded down to ${\sim0.45M_\odot}$, their curves coincide with the overlapping curves of all the models with initial $0.45M_\odot$ RDs, that began with $P\rm_{orb}\sim3.5$ hours. The reason for the WD to be having a negligible influence on determining $P\rm_{orb}$ is that in close RLOF CVs, $R_{RL}\approx R_{RD}$. The results for the models explored in this work yield quite a narrow range of RL radii for the initial models, ranging from ${\sim0.3a}$ to ${\sim0.4a}$\footnote{The entire range of $R_{RL}$ for $0<q<\infty$ is $0<R_{RL}<0.82$ \cite[]{Eggleton1983} implying that most $R_{RL}$ in RLOF CVs would not be far from the above mentioned narrow range.}. This range is a result of the varying $q$ ($M_{RD}/M_{WD}$) which determines the relation between the separation and the RL radius via \cite[e.g.,][eq.2]{Eggleton1983}:
\begin{equation}
\frac{R_{RL}}{a}=\frac{0.49q^{\frac{2}{3}}}{0.6q{\frac{2}{3}}+{\rm{ln}}(1+q^{\frac{1}{3}})}
\end{equation}
which is the only way the WD has an effect on determining the separation. This is apparent in the right panel of Figure \ref{fig:MRD_a_Porb} as the spaces between curves with different WD masses. These spaces are (nearly) absent in the orbital period curves, because of Kepler's third law which dictates the relation $P{\rm_{orb}}\propto(a^3/(M_{WD}+M_{RD}))^{1/2}$ thus compensating for the differences in the WD mass and eliminating almost entirely the dependence of the orbital period on the WD mass. 
All seven of the $P\rm_{orb}$ curves continue to decrease, in identical fashion, until reaching the minimum orbital period of $\sim1.4$ hours for a RD mass of $\lesssim0.15M_\odot$. From this point $P\rm_{orb}$ remains fairly constant until the end of the simulation, at $M_{RD}\lesssim0.1M_\odot$. The minimal period is the result of the transition of the RD donor from a star to a sub-stellar object. For a RD with a mass as low as $\lesssim0.15M_\odot$, hydrogen burning starts to become less efficient, approaching the brown dwarf limit. For the stellar mass-radius relation, $R\propto M^\alpha$, $\alpha$ begins to transition from order $\sim1$, which is characteristic of main sequence stars, to $\sim-\dfrac{1}{3}$, which is characteristic of degenerate objects. This means that for these extremely low mass RDs, the radius does not contract with mass loss as efficiently as before, halting the decrease in the orbital period \cite[]{Howell2001,King2002,Knigge2011a,Knigge2011}. 

\begin{figure}
	\begin{center}
		{\includegraphics[trim={0.7cm 1.0cm 1.2cm 0.1cm},clip ,
			width=1.0\columnwidth]{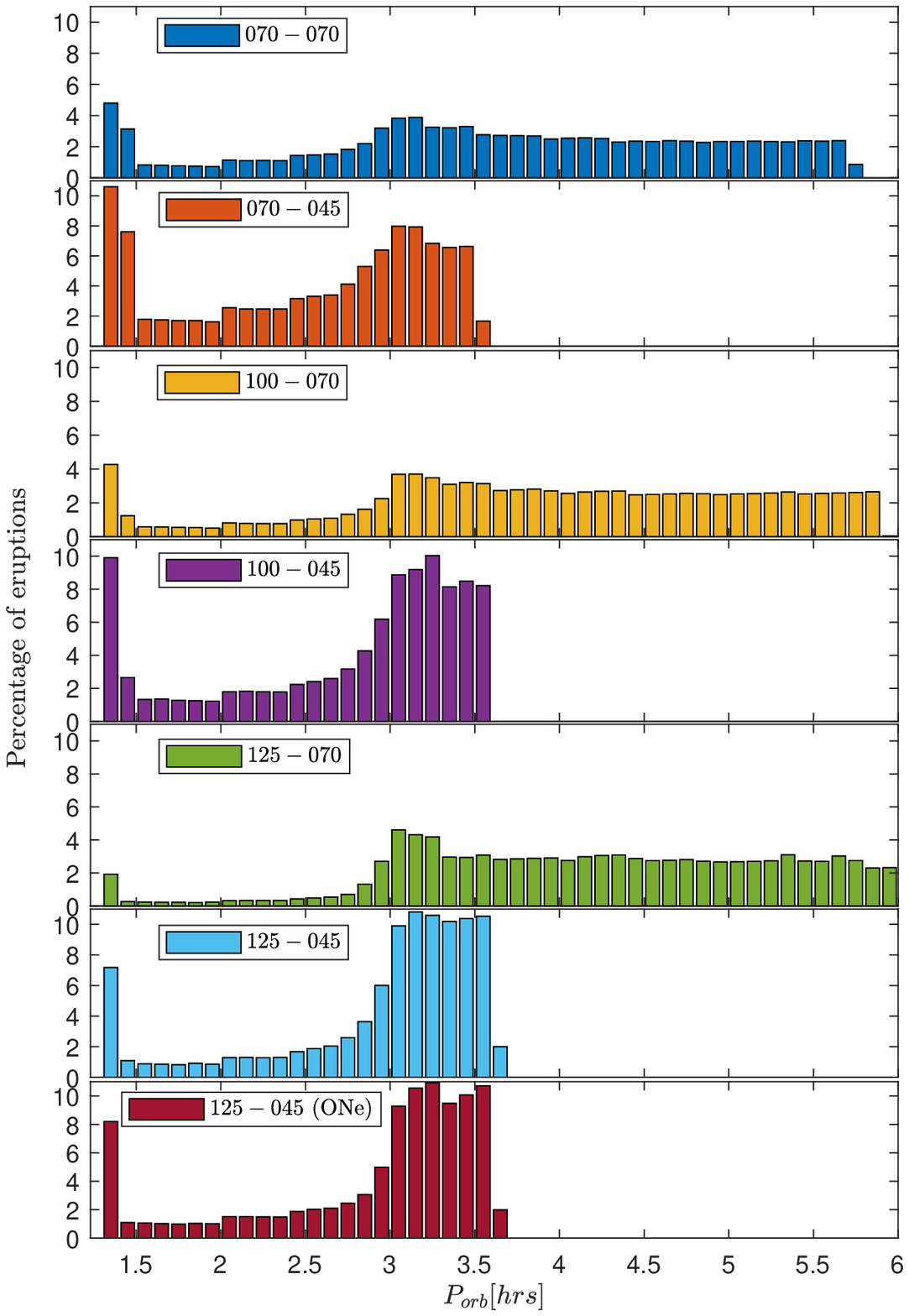}}
		\caption{Percentage of eruptions vs. orbital period. \label{fig:Erup_Porb}}
	\end{center}
\end{figure}

Although the relation between the RD mass and the orbital period is almost linear for almost the entire evolution, it does not reflect the detection probability of systems with different orbital periods. As discussed earlier and shown in Figure \ref{fig:trec} (see right panel), after the MB turns off, the time between eruptions ($t\rm_{rec}$) increases drastically. Since this type of system is only detected during eruption, systems with long $t\rm_{rec}$ --- i.e., systems that rarely erupt --- will rarely be detected. These systems comprise of RDs below the MB turn-off, i.e., with masses below $\sim0.35M_\odot$ which, as seen in Figure \ref{fig:MRD_a_Porb}, is correlated with orbital periods below $\sim3$ hours. Figure \ref{fig:Erup_Porb} shows the percentage of eruptions (total number of eruptions that occurred within each $P\_{orb}$ bin divided by the total evolutionary number of eruption for each model) that occur within each $P\_{orb}$ bin, exhibiting a dip in the percentage of eruptions in the regime of roughly $\sim2-3$ hours (see also figures 3 and 4 in \cite{Hillman2020}). This is in agreement with the period gap, within which a relatively small number of systems have been discovered \cite[e.g.,][]{Knigge2011}.

\begin{figure}
	\begin{center}
		{\includegraphics[trim={0.5cm 0.0cm 0.0cm 0.3cm},clip ,
			width=1.0\columnwidth]{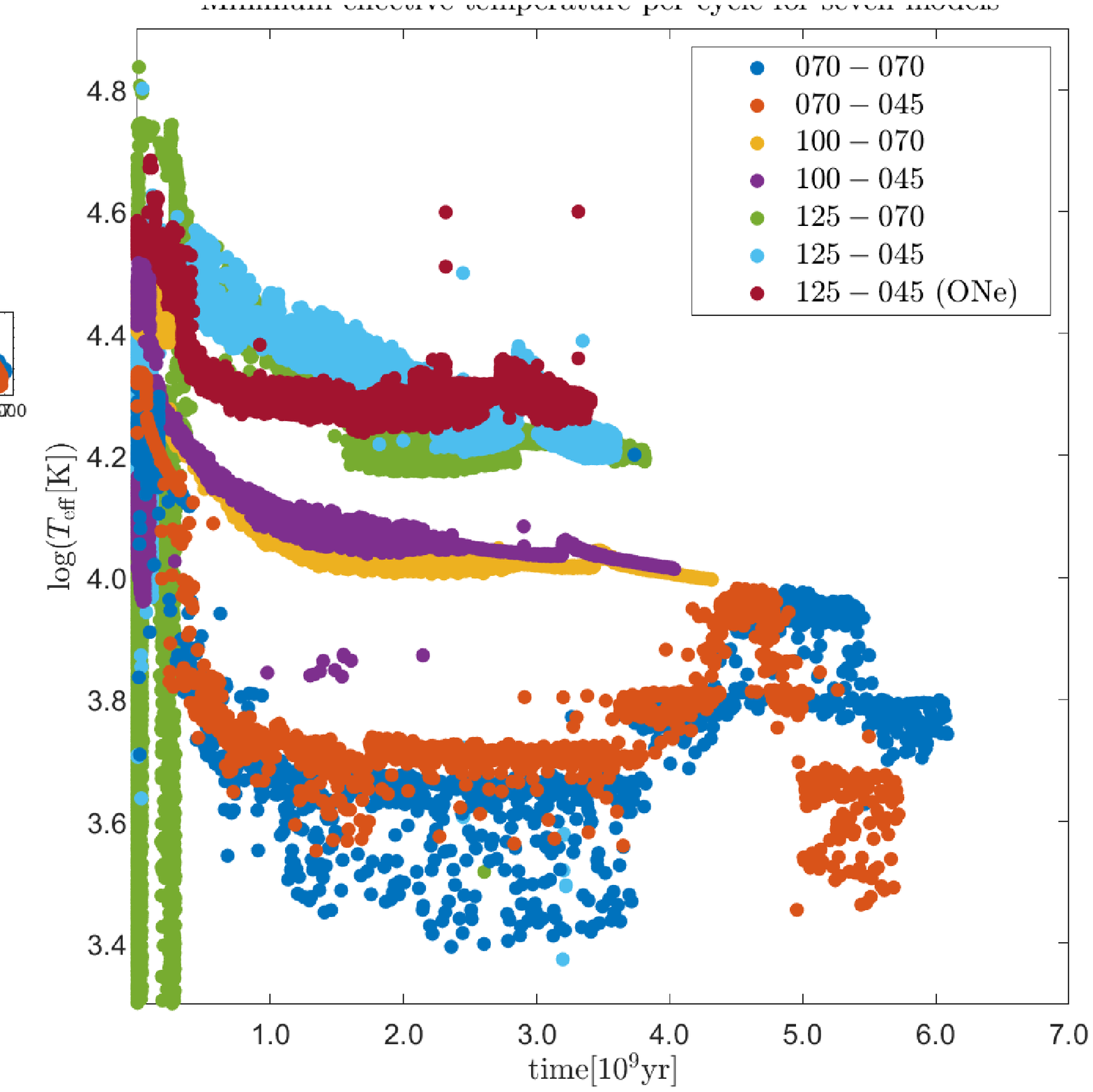}}
		\caption{\label{fig:Teff}Minimal WD effective temperature for the seven models.}
	\end{center}
\end{figure}

In addition,  Figure \ref{fig:Teff} presents the minimal WD effective temperature ($T\rm_{eff}$) per cycle, showing that its behavior follows the behavior of the accretion rate --- higher for higher accretion rates, and therefore, for less accreted mass and vice versa. This is because lower accretion rates produce stronger TNRs, causing more of an expansion of the outer layers of the WD than for faster accretion rates. When the outer layers expand, the effective temperature decreases, which means that lower accretion rates result in lower minimum effective temperatures (for a given WD mass). 
For the same reason, lower accretion rates lead to the ejection of more mass than do higher accretion rates. The critical $m\rm_{acc}$ takes more time to accumulate, thus allowing more time to mix deeper into the envelope \cite[]{Prialnik1984,Prialnik1986,Iben1992}, resulting in a stronger TNR and a higher ratio $m{\rm_{ej}}/m\rm_{acc}$. This ratio is shown in Figure \ref{fig:mej_macc} (right) where all the models have a substantially higher ratio $m{\rm_{ej}}/m\rm_{acc}$ for the evolutionary epoch of $M_{RD}\gtrsim0.35M_\odot$ than for the evolutionary epoch of $M_{RD}\lesssim0.35M_\odot$, which is explained as follows.

\begin{figure}
	\begin{center}
		{\includegraphics[trim={0.0cm 0.0cm 0.0cm 0.0cm},clip ,
			width=1.0\columnwidth]{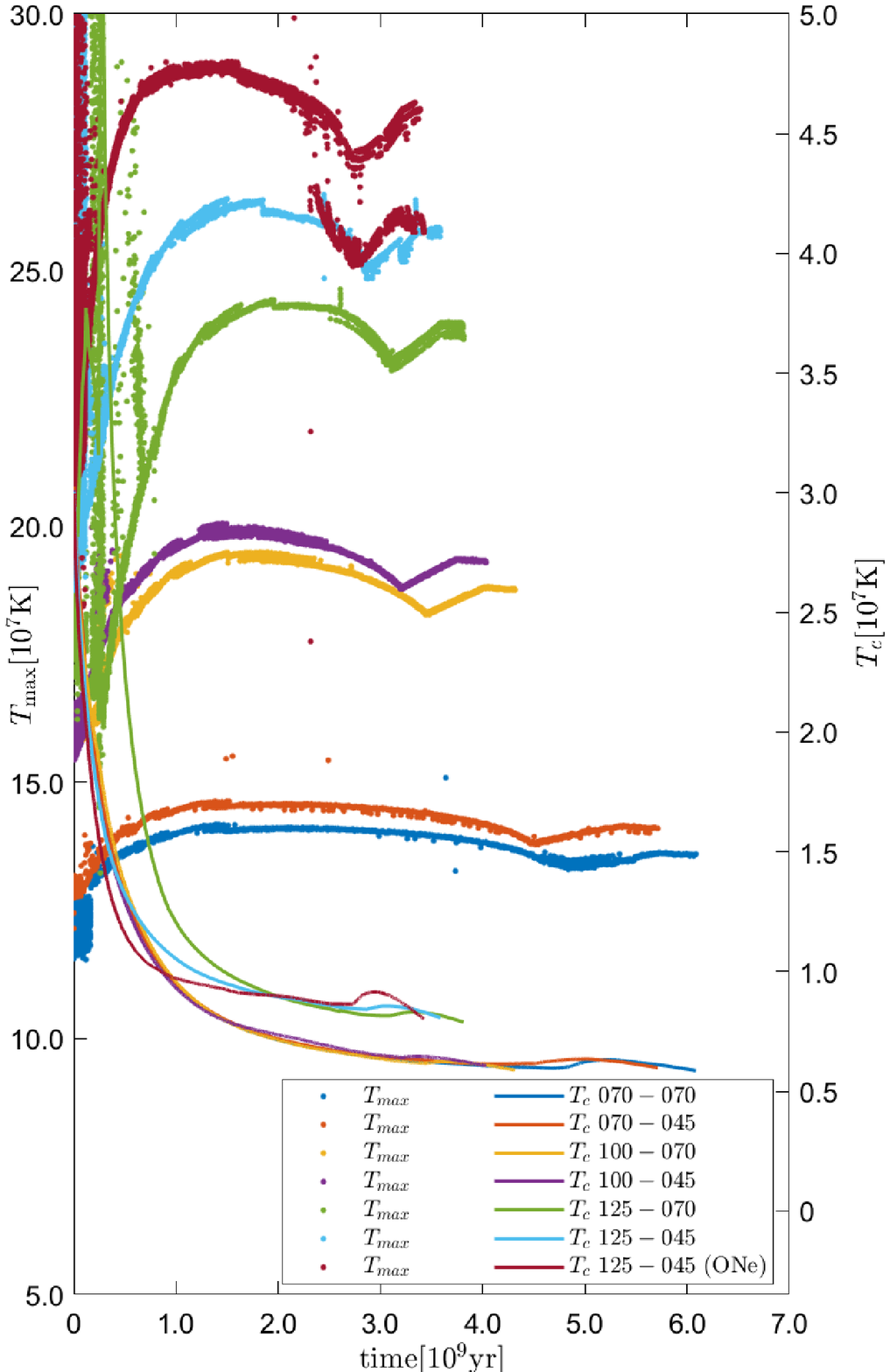}}
		\caption{\label{fig:TcTmax}WD temperature for the seven models. Left axis: maximum envelope temperature ($T{\rm_{max}}$); Right axis: core temperature ($T_c$).}
	\end{center}
\end{figure}

As presented in Figure \ref{fig:Mdot_avg_bfr_aftr}, the accretion rate for $M_{RD}\gtrsim0.35M_\odot$ is of order $\sim10^{-9}-10^{-8}M_\odot yr^{-1}$ and for $M_{RD}\lesssim0.35M_\odot$ it is of order $\sim10^{11}-10^{-10}M_\odot yr^{-1}$. Examining models from \cite{Yaron2005} which are closest to the models presented here (i.e., $M_{WD}=$0.65, 1.0 and 1.25$M_\odot$ for $\dot{M}=10^{-8}$, $10^{-9}$, $10^{-10}$ and $10^{-11}M_\odot yr^{-1}$) and comparing the ratio $m{\rm_{ej}}/m\rm_{acc}$ between their models, yields that for identical WD mass and accretion rate, the ratio $m{\rm_{ej}}/m\rm_{acc}$ depends on the core temperature ($T_c$) of the WD. For $T_c=3\times10^7$[K] their ratio $m{\rm_{ej}}/m\rm_{acc}$ is higher than for $T_c=1\times10^7$[K]. 
The $T_c$ of all the models was followed throughout evolution and presented in Figure \ref{fig:TcTmax}, showing that it begins at $3\times10^7$[K] and declines in the same manner as the ratio $m{\rm_{ej}}/m\rm_{acc}$, reaching $<1\times10^7$[K] at around the same time that the ratio $m{\rm_{ej}}/m\rm_{acc}$ declines from its high value and stabilizes at roughly $1.1$, consistent with the behavior of the models from \cite{Yaron2005}.

Figure \ref{fig:TcTmax} also shows the maximal WD temperature ($T\rm_{max}$) per cycle, which is correlated with the accreted mass. As explained earlier, more accreted mass means that there was more time for accretion (due to a lower accretion rate). This allows more time for heavy elements (mainly carbon) to be dredged up from the core into the envelope, thus raising the burning temperature and leading to a more powerful TNR. This is exhibited as increases and decreases in $T\rm_{max}$, following the trend of the accreted mass. Although, the more substantial difference in $T\rm_{max}$ exhibited in Figure \ref{fig:TcTmax} is between models of different WD masses --- the $T\rm_{max}$ for models with $M_{WD}=0.70M_\odot$ reaches at most only $\sim1.4\times10^8$[K] while for models with $M_{WD}=1.25M_\odot$ the  $T\rm_{max}$ reaches as high as $\sim2.5\times10^8$[K]. 

\begin{figure*}
	\begin{center}
		{\includegraphics[trim={0.6cm 1.6cm 0.37cm 0.7cm},clip ,
			width=1.0\columnwidth]{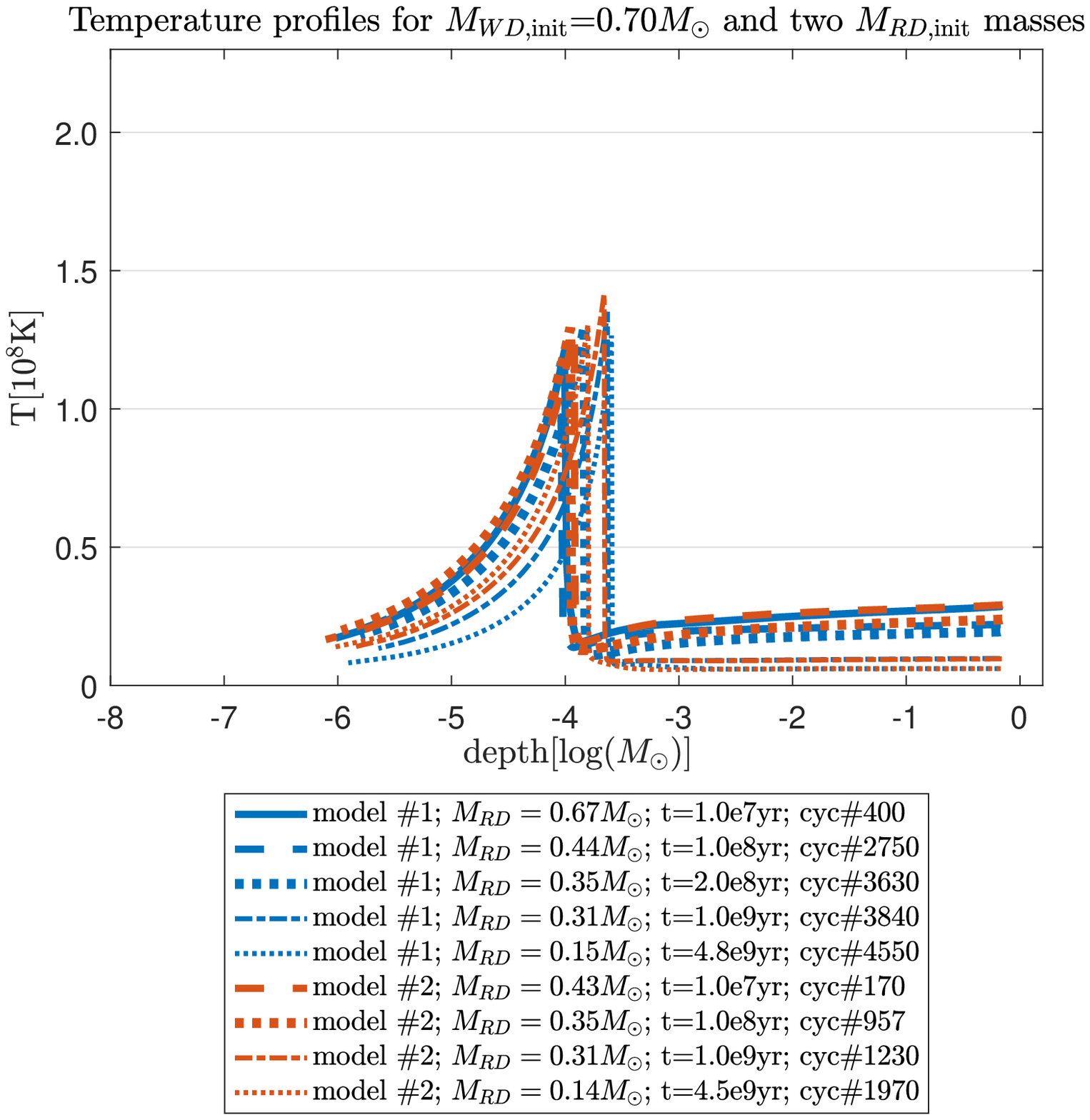}}
		{\includegraphics[trim={0.6cm 1.6cm 0.37cm 0.7cm},clip ,
			width=1.0\columnwidth]{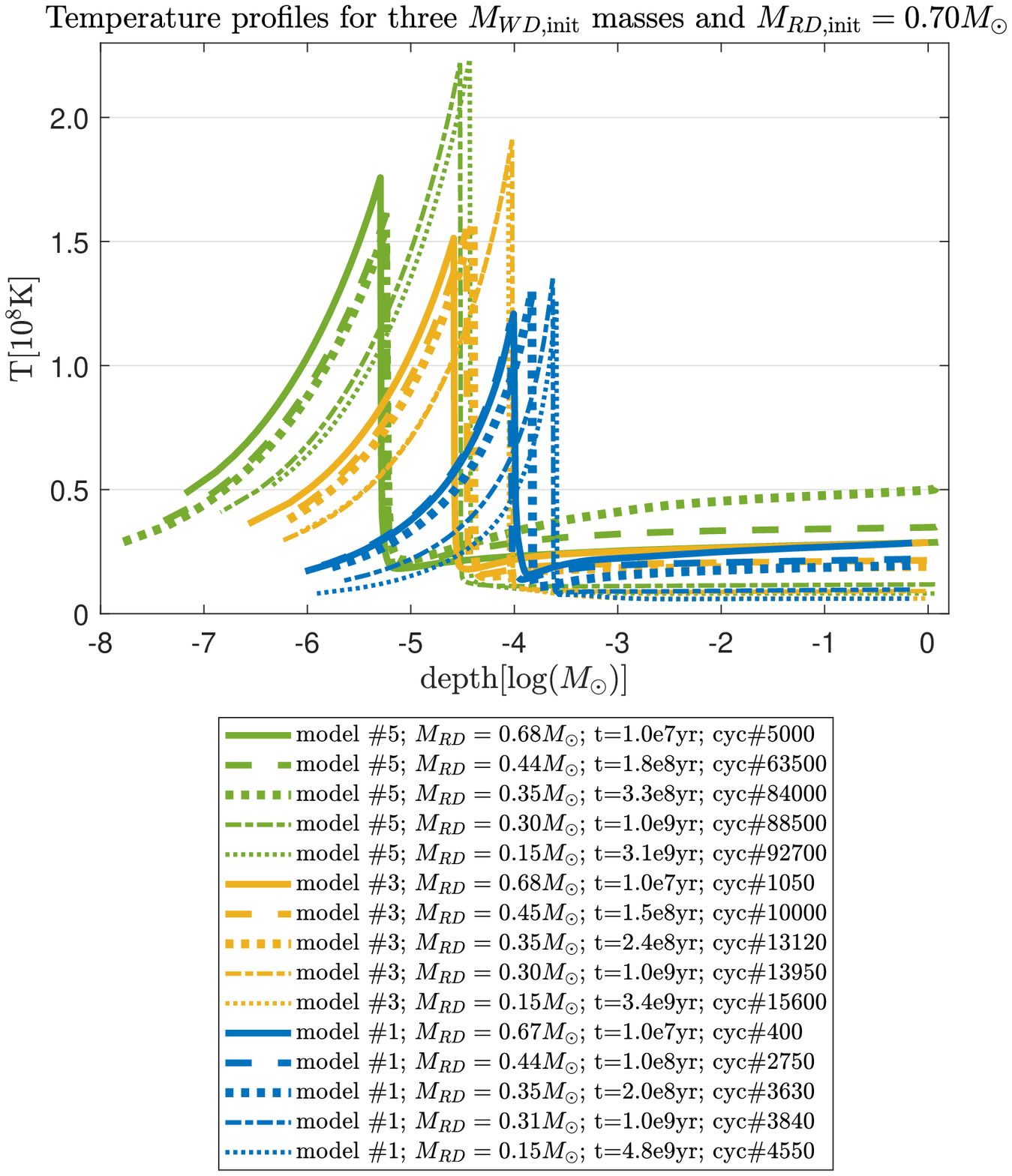}}
		\caption{\label{fig:TempProf}WD temperature vs. depth at several points in time of the evolution for the two models with a common initial WD mass of $0.7M_\odot$ (left) and for the three models with a common initial RD mass of $0.7M_\odot$ (right).}
	\end{center}
\end{figure*}

To further demonstrate the independence of the WD temperature evolution from the RD mass, Figure \ref{fig:TempProf} shows depth profiles of the WD temperature at various points in time throughout evolution. The left panel shows the temperature evolution of the 0.70$M_\odot$ WDs for models with different initial RD masses (models $\#1$ and $\#2$), showing that the curves of the two models coincide and for both models there is a general cooling of the core over time (which is a feature that is evident from Figure \ref{fig:TcTmax} as well). In contrast, the right panel of Figure \ref{fig:TempProf} shows the temperature evolution of the different WD masses in the models with an initial RD mass of 0.7$M_\odot$ (models $\#1$, $\#3$ and $\#5$), all exhibiting a general cooling, but each model developing differently. The less massive WDs have a deeper envelope, a lower maximal temperature ($T{\rm_{max}}$) and less of a change in $T{\rm_{max}}$ throughout evolution.  

\begin{figure}
	\begin{center}
		{\includegraphics[trim={0.0cm 0.0cm 0.0cm 0.0cm},clip ,
			width=1.0\columnwidth]{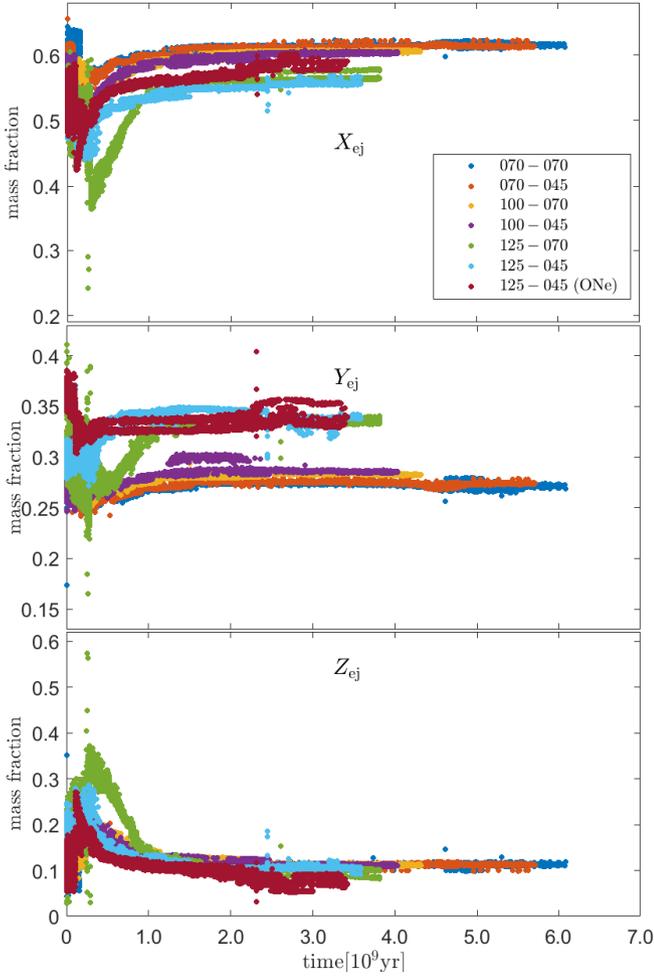}}
		\caption{\label{fig:Xej_Yej_Zej}Mass fractions of hydrogen (top), helium (middle) and heavy element (bottom) abundances in the ejecta for the seven models.}
	\end{center}
\end{figure}

\begin{subfigures}\label{fig:Abundances}
\begin{figure}
	\begin{center}
		{\includegraphics[trim={0.0cm 0.0cm 0.0cm 0.0cm},clip ,
			width=1.0\columnwidth]{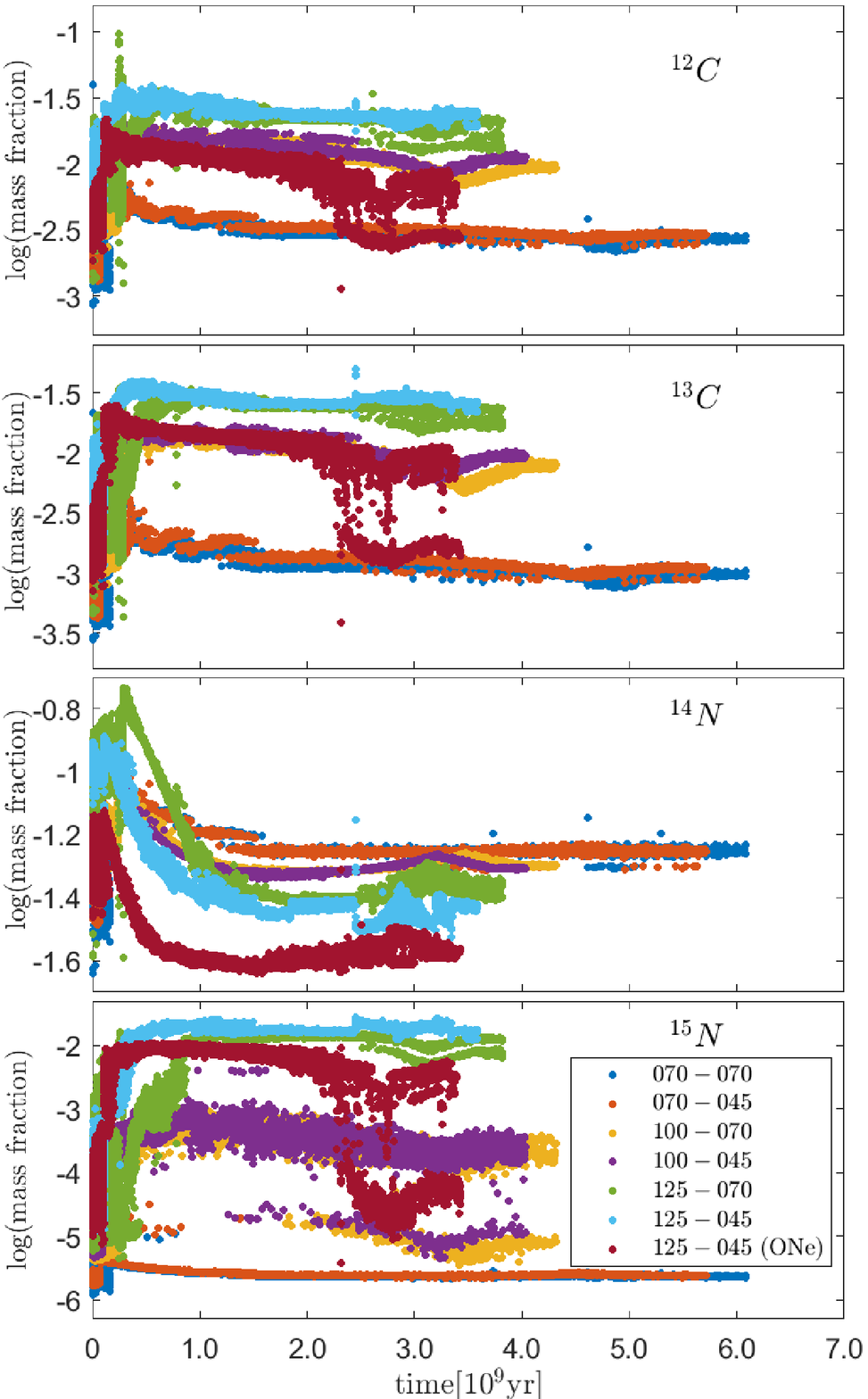}}
		\caption{\label{fig:Abundances1}Heavy element abundances in the ejecta.}
	\end{center}
\end{figure}
\begin{figure}
	\begin{center}
		{\includegraphics[trim={0.0cm 0.0cm 0.0cm 0.0cm},clip ,
			width=1.0\columnwidth]{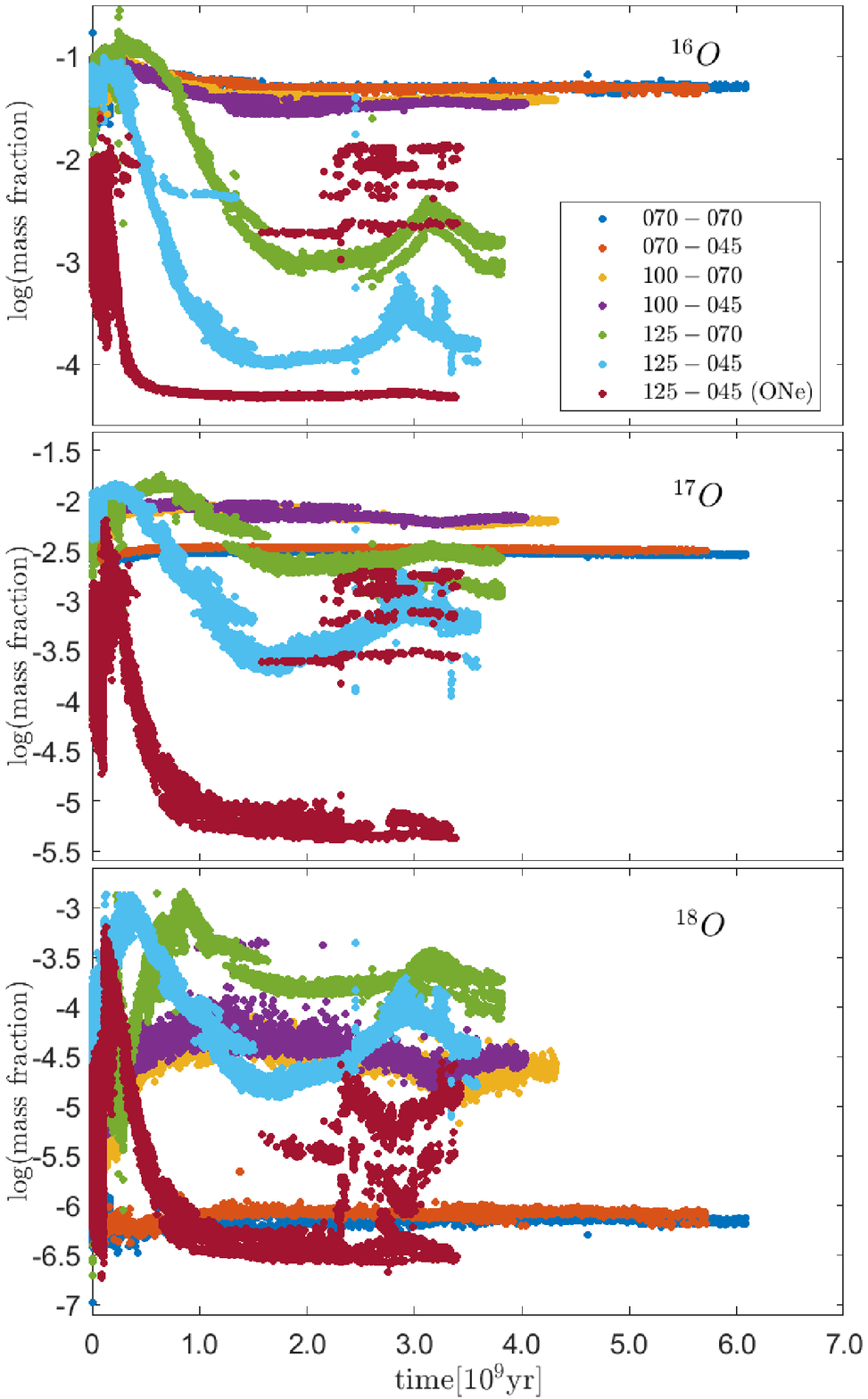}}
		\caption{\label{fig:Abundances2}Heavy element abundances in the ejecta --- cont.}
	\end{center}
\end{figure}
\begin{figure}
	\begin{center}
		{\includegraphics[trim={0.0cm 0.0cm 0.45cm 0.1cm},clip ,
			width=1.0\columnwidth]{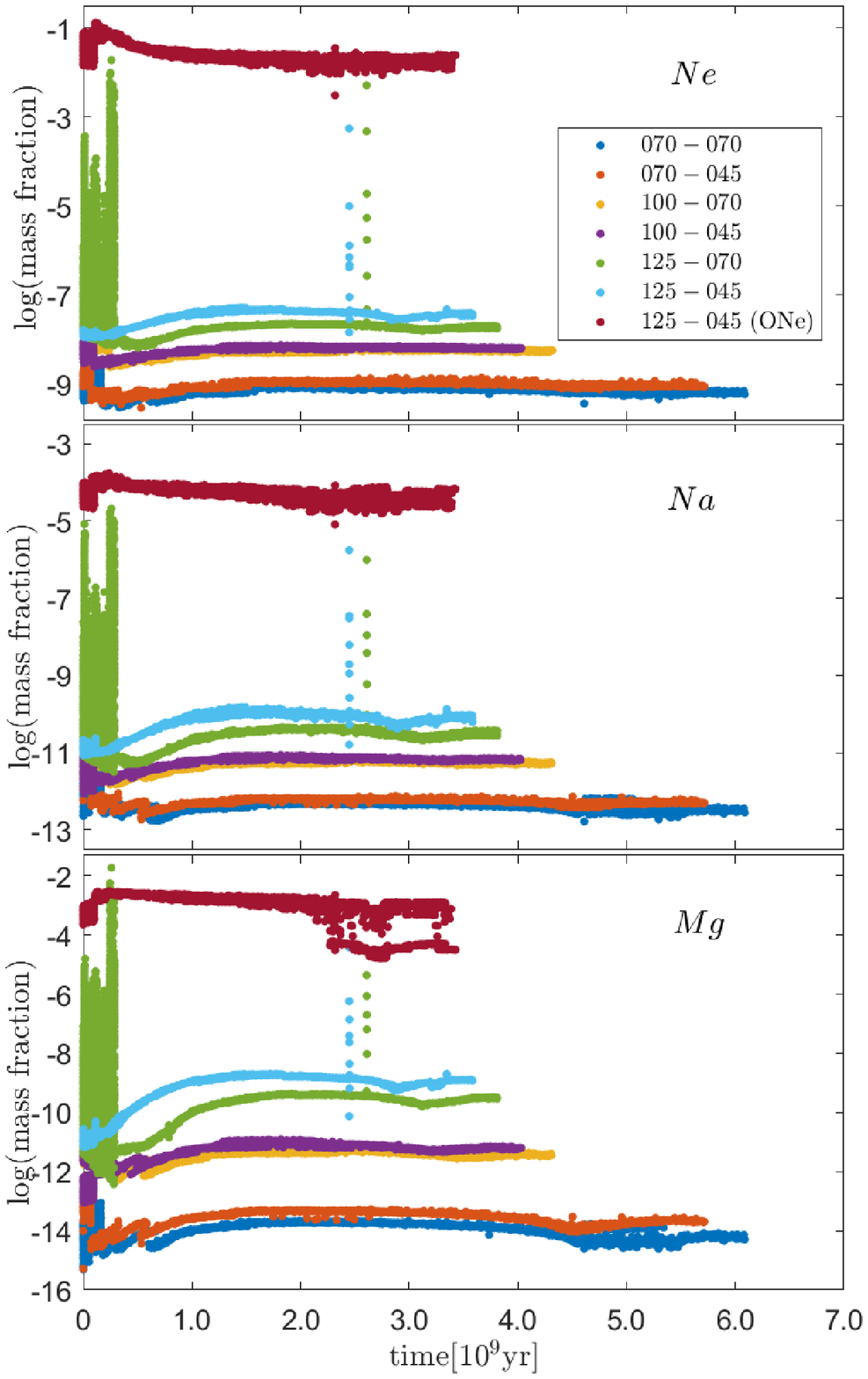}}
		\caption{\label{fig:Abundances3}Heavy element abundances in the ejecta --- cont.}
	\end{center}
\end{figure}
\begin{figure}
	\begin{center}
		{\includegraphics[trim={0.0cm 0.0cm 0.0cm 0.0cm},clip ,
			width=1.0\columnwidth]{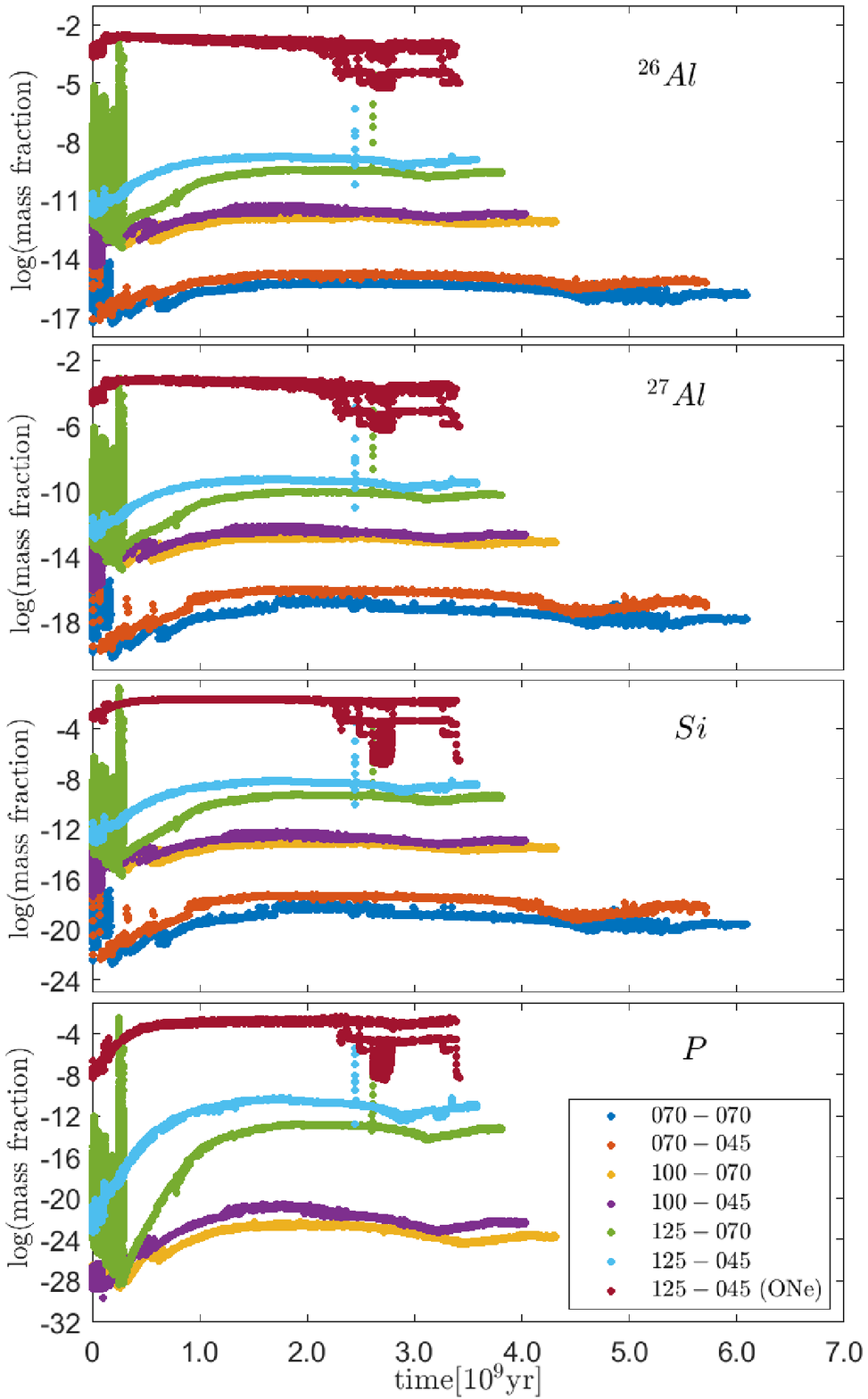}}
		\caption{\label{fig:Abundances4}Heavy element abundances in the ejecta --- cont.}
	\end{center}
\end{figure}
\end{subfigures}

As shown earlier, the ratio of ejected to accreted mass, demonstrates how in certain perspectives, the differences between models are small, while the evolutionary changes are substantial. 
The ejecta composition shares some of these common features, while emphasizing some of the differences. Figure \ref{fig:Xej_Yej_Zej} shows the hydrogen ($X\rm_{ej}$) helium ($Y\rm_{ej}$) and heavy element ($Z\rm_{ej}$) abundances (${X\rm_{ej}+Y\rm_{ej}+Z\rm_{ej}=1}$) in the ejecta per eruption. All the curves in all three panels behave like the ratio $m{\rm_{ej}}/m\rm_{acc}$ (Figure \ref{fig:mej_macc}, right) --- a sharp difference between the early behavior and the behavior for the rest of the evolution --- when the ratio $m{\rm_{ej}}/m\rm_{acc}$ is high, the heavy element abundance in the ejecta ($Z\rm_{ej}$) is high as well. $Z\rm_{ej}$ is essentially identical for all seven models, with the curves overlapping each other. The difference arises in the helium abundance. The models with $M_{WD}=1.25M_\odot yr^{-1}$ have a higher mass fraction of helium in the ejecta --- $\sim34\%$ (and therefore a correspondingly lower mass fraction of hydrogen) as opposed to $\sim26-28\%$ for the other less massive WD models (for which the helium mass fraction for models with $M_{WD}=1.0M_\odot$ is slightly higher than the helium mass fraction for models with $M_{WD}=0.7M_\odot$). This is a direct consequence of the maximum temperature during eruption, which is higher for the more massive WDs, and therefore the hydrogen is burnt into helium more efficiently. 

The enrichment of heavy elements in the ejecta is mostly due to dredge up of carbon from mixing, increasing the energy output from the CNO cycle. This means that a longer accretion phase --- which (for a given WD mass) is the result of a lower accretion rate --- will yield a higher enrichment of heavy elements in the ejecta. This is in agreement with other authors \cite[e.g.,][]{Prialnik1984,Iben1992,Prikov1995,Yaron2005}. 
Figure \ref{fig:Abundances} displays the breakdown of the heavy element ejecta abundances, showing that the WD mass has a significant effect. The ejecta in models with $M_{WD}=0.7M_\odot$ constitutes only ${\sim0.3\%}$ of carbon ($^{12}C$) and ${\sim6\%}$ nitrogen ($^{14}N$) (Figure \ref{fig:Abundances1}). The percentage of carbon grows with WD mass while the percentage of nitrogen decreases, exhibiting ${\sim1\%}$ and ${\sim5\%}$ of carbon and nitrogen respectively for $M_{WD}=1.0M_\odot$ and ${\sim2.5\%}$ and ${\sim4-5\%}$ of carbon and nitrogen for $M_{WD}=1.25M_\odot$. This is the result of a (slightly) higher temperature converting carbon and oxygen to nitrogen more efficiently. For this reason, the percentage of oxygen ($^{16}O$) in Figure \ref{fig:Abundances2} also decreases with increasing WD mass, constituting $\sim5\%$ for models with $M_{WD}=0.7M_\odot$, $\sim3.5\%$ for models with $M_{WD}=1.0M_\odot$ and $\sim0.02-0.1\%$ for models with $M_{WD}=1.25M_\odot$. The remaining elements, presented in Figures \ref{fig:Abundances3} and \ref{fig:Abundances4} show trace amounts, and all exhibit the trend of increasing with increasing $M_{WD}$. There is one exception --- the ONe model, which is now discussed.

\section{Oxygen neon core}\label{sec:ONe}
Throughout the previous section, the results of the different models were compared, focusing on the WD and RD masses for which all but one comprise a carbon-oxygen (CO) core. WDs that are more massive than ${\sim1.2M_\odot}$ are commonly accepted as having a core comprised of oxygen and neon (ONe) rather than a CO core due to the nuclear processes that occur in their progenitor main sequence stars \cite[]{Law1983,Gutierrez1996,Gil2001}. Therefore, included in this study is one model ($\#$7) which consists of the same stellar masses as model $\#6$ but with an ONe core. This is for comparison, and determination of which features are affected by the core composition and which are not. One of these features, which is observable, and prominently deviates from the other models, is the ejecta abundance. While the ratio $X{\rm_{ej}}:Y{\rm_{ej}}:Z{\rm_{ej}}$ for this model is essentially the same as for its corresponding CO core model, the breakdown of the heavy elements is entirely different, with less carbon, nitrogen and oxygen. Namely ${\sim1\%}$ of carbon, which is similar to the less massive 1.0$M_\odot$ CO WDs; ${\sim2.5\%}$ of nitrogen, as opposed to ${\sim4\%}$ in the corresponding CO model; and considerably less oxygen as well, ${\sim0.007\%}$ which is about a quarter of the oxygen fraction in the corresponding CO model. On the other hand, all the rest of the elements, which for the CO core model are at the most a mass fraction of order $10^{-8}$, in the ONe core model they constitute the better part of the heavy element composition with ${\sim2.5\%}$ of neon (Ne); ${\sim2\%}$ of silicon (Si); ${\sim0.2\%}$ of magnesium (Mg) and of aluminum 26 ($^{26}$Al); ${\sim0.1\%}$ of phosphorus (P); ${\sim0.06\%}$ of aluminum 27 ($^{27}$Al); and ${\sim0.007\%}$ of sodium (Na). The favoring of neon over carbon is simply due to the WD core containing neon instead of carbon. The reduction of nitrogen and oxygen is due to the decrease in carbon reducing the rate of the CNO process. The enrichment in heavier elements, such as, silicon, phosphorus and magnesium is a result of the WD core containing neon which is required for the production of these elements. Novae have been observed to have ejecta enriched in heavy elements. For example, \cite{Delbourgo1985} report V1500 Cyg and Nova Aql 1982 being enriched in neon, and the enhancement of neon and other heavy elements was reported for V693 CrA, V351 Pup, V1974 Cyg and V838 Her \cite[][and references therein]{Cassatella2002}. \cite{Nomoto1987} suggested that enrichment in neon must be as a result of it being initially present because the CNO cycle does not lead to the efficient production of neon at temperatures typical of nova. \cite{Jose1998} performed a series of simulations of novae on CO and ONe core WDs for different mixing levels. Comparing between two of their models with WDs of $1.15M_\odot$ and identical mixing levels yields a decrease in the abundance of carbon, nitrogen and oxygen in the ejecta and a substantial increase in heavy elements, similar to the results presented here. \cite{Jose2020} perform simulations that combine 1-D and 3-D methodologies, and obtain, for a $1.25M_\odot$ WD accreting at $10^{-10}M_\odot yr^{-1}$, an ejecta that is enriched relative to a CO core. The enrichment obtained here is similar to their results. The increase in heavy element production entails a somewhat higher burning temperature as well, which may be seen in Figure \ref{fig:TcTmax} to be $\sim2.8\times10^8$[K] as opposed to the CO core model that has a maximum WD core temperature of $\sim2.6\times10^8$[K].

The ONe WD core temperature ($T_c$) cools in the same manner as the core of the corresponding CO model. This is because the evolution of $T_c$ is dependent on the time between eruptions \cite[]{Epelstain2007} which is dependent on the amount of required accreted mass ($m{\rm_{acc}}$). As shown earlier, $m{\rm_{acc}}$ is determined by the WD mass ($M_{WD}$) and accretion rate ($\dot{M}$), which behave in a similar manner as the corresponding CO model. Since the WD effective temperature ($T{\rm_{eff}}$) (Figure \ref{fig:Teff}) is also dependent on $M_{WD}$ and $\dot{M}$ \cite[]{Epelstain2007}, it behaves in the same manner as the CO core model as well. In fact, the only feature that shows a significant deviation from the CO core model is the ejecta abundance, as described above. This is in agreement with \cite{Yaron2005}, who explain that the mechanism leading to a nova eruption is the result of a TNR under electron degeneracy conditions, which is determined by the total mass fraction of heavy elements. Thus, replacing carbon with neon does not alter the required conditions, so it does not alter the amount of required $m\rm_{acc}$ and therefore does not entail significant changes to eruptive features other than the ejecta abundances. 

\section{Conclusions}\label{sec:conclusions}
Using the self-consistent, feedback dominated combined code for evolving a nova producing binary system, while systematically changing one of the binary masses at a time, has led to the primary conclusion that all seven of the models presented here have the same general evolutionary behavior. Despite the common general behavior, there are certain features that vary between models and are distinctly determined by the WD mass while the RD mass has a negligible contribution and vice versa, as detailed below.

\subsection*{Common features}

\begin{itemize}
	\item Evolutionary time - All of the models complete their evolution from first Roche-lobe contact of the RD donor until the donor is eroded down to $<0.1M_\odot$ on a time scale of a few Gyr. This is regardless of the initial RD mass, even though the difference between the two RD masses used in this study is almost a factor of two. This is because only a small fraction of the total evolutionary time is spent reducing the RD mass to $0.35M_\odot$, thus the total evolutionary time is essentially the time required for the WD to erode a $0.35M_\odot$ RD donor.
	
	\item Accretion rate - During a single accretion phase, $\dot{M}$ for all of the models decreases rapidly and then increases slowly. In addition, on a secular time scale, for all of the models, the initial accretion rate per cycle decreases throughout evolution. This demonstrates that the mass transfer rate from a donor to a WD in a RLOF CV is far from being constant. Observations of two different systems with the same WD masses, RD masses and separation may exhibit entirely different behaviors (NL, DN, CN or detached RD-WD binary) if the observations "catch" the systems at different points in time of a nova cycle, exhibiting very different mass transfer rates.
\end{itemize}

\subsection*{Features determined by the RD}
\begin{itemize}
\item Accretion rate - For RD masses above the MB threshold the average accretion rate is about 50$\%$ higher for \textit{initially} more massive RDs. Nevertheless, in general the average accretion rate for all of the models when ${M_{RD}\gtrsim0.35M_\odot}$ is of order $\sim10^{-9}M_\odot yr^{-1}$ while it is of order only several times $10^{-11}M_\odot {yr^{-1}}$ for $M_{RD}\lesssim0.35M_\odot$. This is because of the MB threshold occurring at $\sim0.35M_\odot$ and may serve as a tool in setting a higher or lower limit on the donor mass of observed systems provided there are observationally estimated accretion rates. 

\item Ejecta - The total Z of the ejecta indicates whether the mass of the RD is higher or lower than 0.35$M_\odot$, being of order $\sim35\%$ for cases with a RD mass above the MB threshold, and only $\sim15\%$ for cases with a RD mass below it. This is because the low accretion rate that characterizes RDs below the MB threshold allows more time for mixing of the accreted matter with the core.

\end{itemize}	

\subsection*{Features determined by the WD}
\begin{itemize}
\item Evolutionary time - Although all the models erode their donor on the same time \textit{scale}, the time is shorter for more massive WDs. In fact, this study finds that the erosion time is roughly inversely proportional to the WD mass ($M_{WD,1}/M_{WD,2}\propto t_2/t_1$) e.g., doubling the mass of the WD would reduce the total evolutionary time by half. 
	
\item Accretion rate - For the $0.70M_\odot$ and $1.0M_\odot$ models with donors below $0.35M_\odot$ the accretion rate during a cycle (and secularly over many cycles), decreases to a state of essentially no mass transfer for periods of many millennia, until the separation decreases sufficiently to resume mass transfer. This supports the hibernation theory \cite[]{Shara1986} showing that in between nova eruptions the system has a long detached epoch. These type of epochs may be seen for the more massive $1.25M_\odot$ WD models as well, however they may occur only when the donor is of a very low mass, thus, statistically a hibernating system with a $\sim1.25M_\odot$ WD would be very rare.

\item Ejecta: He - The helium mass fraction in the ejecta of the models with a $1.25M_\odot$ WD is of order $\sim35\%$ whereas the models with lower mass WDs have a lower mass fraction of helium. This means that the mass fraction of the helium in the ejecta may assist in estimating the the WD mass. 

\item Ejecta: CNO - The CNO abundance in the ejecta is informative of the WD mass as well.
The models with a WD mass of $0.70M_\odot$ have $\sim20$ times more nitrogen (14N) than carbon (12C), while the models with a $1.0M_\odot$ WD have $\sim5$ times more and the models with a $1.25M_\odot$ WD have only twice more 12C than 14N. This means that the ratio 12C/14N in the abundance of an observed ejected nova shell may be indicative of the WD mass. The oxygen (16O) abundance refines this by being similar to the 14N abundance for the models with a $0.70M_\odot$ WD, more than half the 14N abundance for the models with a $1.0M_\odot$ WD and only a few percent the 14N abundance for the models with a $1.25M_\odot$ WD. These are observable measures which can assist in estimating the WD mass. For instance, the ratio of carbon to nitrogen observed in nova V1370 Aql\footnote{Aql 1982} is $\sim0.28$ and the ratio of oxygen to nitrogen is $\sim0.31$ \cite[]{Vanlandingham1996}. Comparison with the abundances obtained here places the WD mass between $1.0M_\odot$ and $1.25M_\odot$ which is in agreement with \cite{Shara2018} who found the WD mass of V1370 Aql via modeling to be $\sim1.13M_\odot$ based on observed amplitude and decline time of the light curve. 

\end{itemize}

\subsection*{ONe composition}
The evolution of the ONe model (model $\#7$) shows results which are almost identical to those of the CO model $\#6$ except for the breakdown of the heavy element abundances, displaying considerably less oxygen and non-negligible amounts of heavy elements, such as, neon, silicon and magnesium which are comparable with the mass fractions of carbon and nitrogen. Observing these high abundances in the ejecta of a nova would indicate a WD with an ONe core, which is indeed observed for novae with estimated high mass WDs such as, V838 Her\footnote{Her 91} which shows a high C/O ratio and neon and sulfur enrichment \cite[]{Vanlandingham1996}, 
and QU Vul \cite[]{Mikolajewska2017}, V693 CrA \cite[]{Williams1985} and V382 Vel \cite[]{Shore2003} (among others) which exhibit a strong presence of magnesium, neon, silicon and aluminum in their spectra \cite[]{Mikolajewska2017}. These enrichments are not observed in CO novae such as the seven specified in \cite{Cassatella2005} and are not present in the results of the CO nova models studied here.

\subsection*{Evolutionary age}
An important feature that can not be deduced from the masses of the binary system is the evolutionary age of an observed system. Since the rate that the RD loses mass is the accretion rate, which can be expressed by the average value per cycle as $m{\rm_{acc}}/t{\rm_{rec}}$, and the WD loses mass at an average rate equal to $(m{\rm_{ej}}-m{\rm_{acc}})/t{\rm_{rec}}$, and $m{\rm_{ej}}\simeq m{\rm_{acc}}$, therefore over the entire evolutionary time needed to erode the RD, the WD loses very little mass and may be considered constant --- as found in this study for all seven models. This means that the evolution of the binary is dictated solely by the WDs mass and that all binaries with a certain WD mass will follow the same path, regardless of the initial RD mass (as explained above). This leads to the conclusion that observationally determining the binary masses of a certain system is \textit{not} an indication of the system's age, since for a given WD mass, a system "born" with a certain RD mass will behave in the \textit{same} manner as a system that was "born" with a more massive RD and has evolved to the same RD mass.

\subsection*{}
The results presented here show that WDs of mass up to about $1.25M_\odot$ with RD companions of initial mass up to about $0.7M_\odot$ will not grow in mass at all. To the contrary, all such models' WDs lose mass monotonically throughout their evolution. Therefore, only initially more massive stellar combinations may determine whether or not a WD in a WD+RD system may accumulate enough mass to sufficiently grow and reach the Chandrasekhar mass of $\sim1.4M_\odot$ that is required for igniting the WD core, and may help in answering the long-standing question of the possibility of WD+RD binary system being a type Ia supernova progenitor.

\appendix
\setcounter{section}{-1}

\section{Implications of the mass loss rate on the RD}
\renewcommand{\thesection}{A}
The code that was developed for the binary evolution presented here uses models of ZAMS donors that do not lose mass. As a result of this assumption, there are two relevant effects of mass loss from a main sequence star that were not included in the simulations. Their implications are discussed below.

\subsection*{Convection zones}\label{sec:app_conv}
\begin{figure}
	\begin{center}
		{\includegraphics[trim={0.0cm 0.0cm 0.0cm 0.0cm},clip ,
			width=1.0\columnwidth]{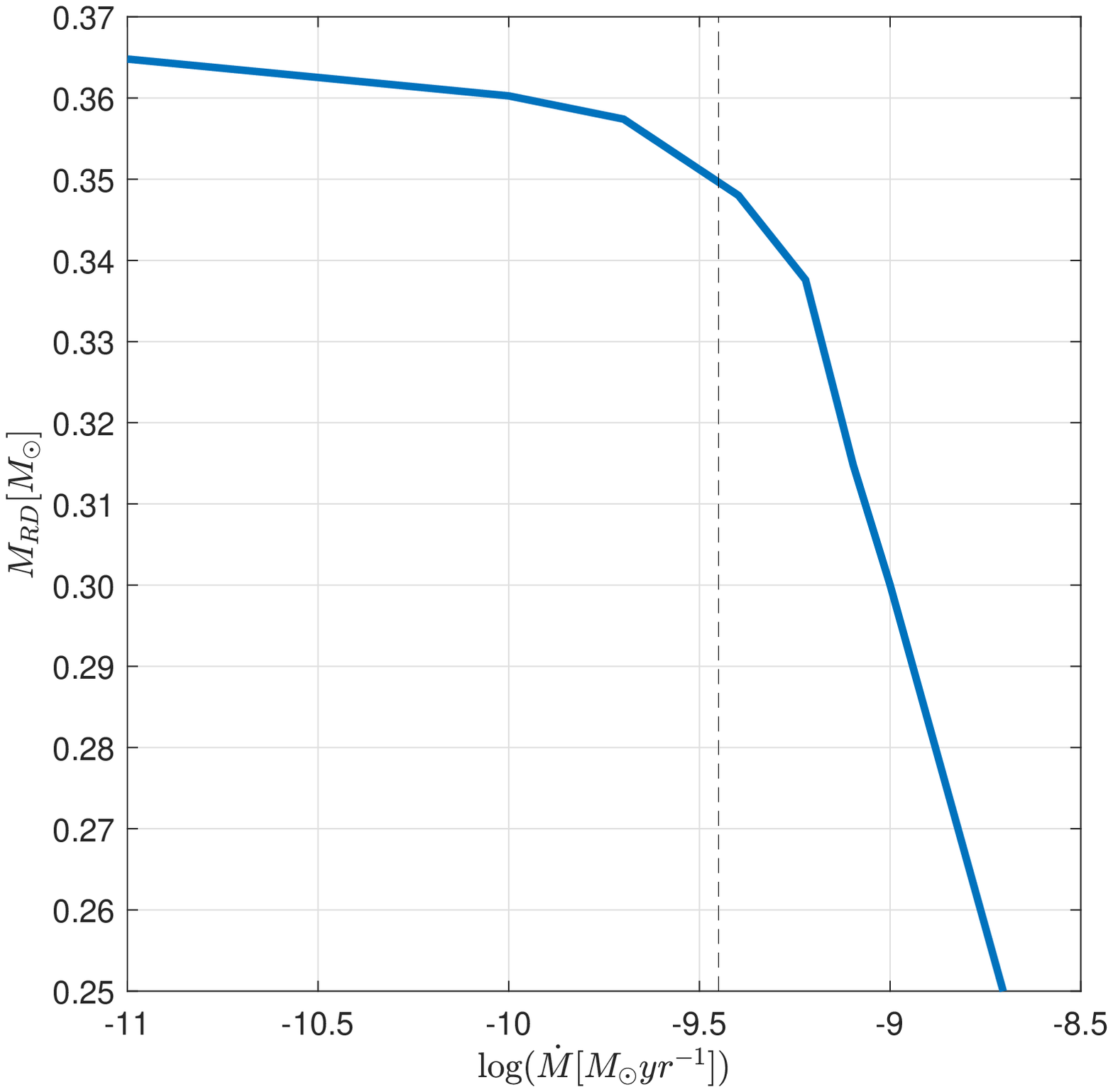}}
		\caption{\label{fig:fully_convective}RD mass vs. the mass loss rate for which it becomes fully convective.}
	\end{center}
\end{figure}

\begin{figure}
	\begin{center}
		{\includegraphics[trim={0.0cm 0.0cm 0.0cm 0.0cm},clip ,
			width=1.0\columnwidth]{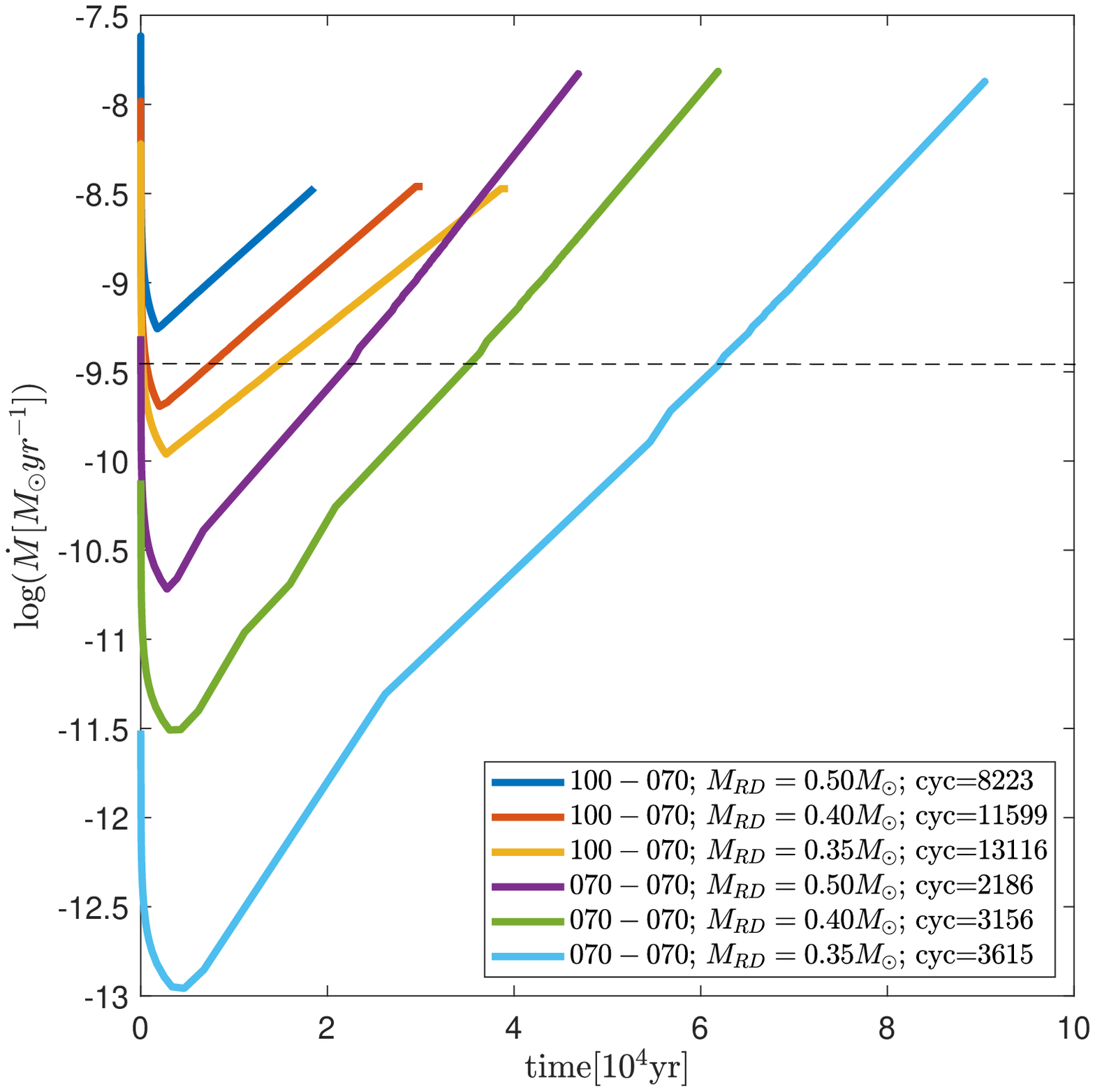}}
		\caption{\label{fig:Mdots_convection_zones}Mass transfer rate vs. time for three cycles corresponding with the RD masses 0.50, 0.40 and 0.35$M_\odot$ each, for models $\#1$ and $\#3$ (see Table \ref{tab:mdls}).}
	\end{center}
\end{figure}

Main sequence (MS) stars with masses of order one solar mass are only partially convective, forming a magnetic field. The presence of this magnetic field in the close binary systems examined in this study is the main source of angular momentum drainage from the system via magnetic braking caused by particles trapped in the magnetic field lines. Lower mass MS stars become fully convective, deeming the magnetic field as redundant. For close binary systems as the ones explored in this work, eliminating magnetic braking (MB) leaves only the relatively weak gravitational radiation (GR) to remove angular momentum. Since the angular momentum loss (AML) due to GR is only about one tenth of the AML due to MB, the RD mass for which the star first becomes fully convective, largely determines at what evolutionary point the accretion rate onto the WD substantially decreases causing the time between eruptions to substantially increase ($\times\sim$10). Therefore, it is important to evaluate the significance of obtaining the precise RD mass for which the star first becomes fully convective. For a RD that is unaffected by its companion and not losing mass the limiting mass under which it is fully convective is of order $\lesssim0.30-0.35M_\odot$ \cite[e.g.,][]{Howell2001} which is in agreement with the value of $\sim0.35M_\odot$ obtained and used in this work. However, in nova producing systems the RD inevitably loses mass yielding the need for an assessment to the affect of mass loss on convection. For this purpose a test --- a series of simulations --- was performed, beginning with RDs of the mass $\sim0.37M_\odot$ --– a mass above which is widely agreed upon that the RD is not fully convective. The simulation was performed multiple times, each time with a different, constant, mass loss rate, until the RD became fully convective. The RD mass at this point was recorded. The results of this exercise are presented in Figure \ref{fig:fully_convective} showing that the limit of roughly $\sim0.35M_\odot$ remains valid as long as the mass loss rate is lower than a few times $10^{-10}M_\odot yr^{-1}$, which is true for more than half of the time between two successive eruptions (furthermore, the limit remains $\gtrsim0.30M_\odot$ for accretion rates as high as $\sim10^{-9}M_\odot yr^{-1}$).  
Now  since the mass loss rate secularly decreases from one cycle to the next (also early in the evolution before the MB ceases), as the evolution progresses, the fraction of the time between two successive eruptions, during which the mass loss rate is low, increases. This is demonstrated in Figure \ref{fig:Mdots_convection_zones} for three cycles each, from two of the models from this work, the three cycles being correlated with the RD masses of 0.5, 0.4 and 0.35$M_\odot$ before the MB is turned off. Examining the three cycles of model $\#1$ (i.e., with the WD mass of 0.7$M_\odot$) shows that although $\dot{M}$ is high while the RD mass is high, as the RD loses mass, the mass loss rate decreases, reaching a state at 0.35$M_\odot$ for which $\dot{M}$ is lower than the limit for MB turn-off for more than half of the cycle accretion time.  Figure \ref{fig:Mdots_convection_zones} additionally shows three cycles, for the same RD masses, from model $\#3$ (i.e., with a WD mass of 1.0$M_\odot$), showing that also for systems with more massive WDs, when the RD is $0.35M_\odot$ the $\dot{M}$ is below the limit for MB turn-off. During these epochs of low $\dot{M}$ (below a few times $10^{-10}M_\odot yr^{-1}$), neglecting the affect of mass loss on convection is certainly justified. For the remainder of the cycle, for which $\dot{M}$ is higher than the limit found, it is unknown to date (and may be worthy of a careful study) if it is possible for a star to revert back to a non-fully convective state, in which case the remainder of the cycle might be shortened to a certain extent, nonetheless, \textit{this does not change the rate of the low $\dot{M}$ nor the length of time that the $\dot{M}$ is low, meaning that it does not change the conclusion that these systems eventually enter hibernation}.

\subsection*{Radial bloating}
A RD losing mass will be slightly out of thermal equilibrium, resulting in the expansion of the RD's atmosphere \cite[e.g.,][]{Knigge2011}. Since the mass loss rate is determined by the amount of RLOF (i.e., $R{\rm_{RD}}-R\rm_{RL}$), bloating of the RD's envelope may cause the mass loss rate to increase, noting that bloating may also cause the density of the envelope to \textit{decrease}, having an opposite affect on the mass transfer rate. Nevertheless, in order estimate the RD bloating, a series of simulations were carried out for different RD masses at different mass loss rates, and the resulting, bloated radius was compared with the radius of a RD of the same mass that is not losing mass --- as used in this work. 
This comparison was performed for RD masses ranging $0.2-0.7M_\odot$ and mass loss rates in the range $10^{-12}-10^{-8}M_\odot yr^{-1}$, 
yielding the conclusion that for RD masses of below roughly $\sim0.3M_\odot$ losing mass at rates higher than a couple times $\sim10^{-9}M_\odot yr^{-1}$ the bloating becomes significant, where in extreme cases it can reach a few percent or more. For the rest --- the majority --- of the regime of RD masses and of mass loss rates, the RD bloating is negligible --- less then $\sim1\permil$ as compared to a RD of the same mass not losing mass.

\subsection*{}
This means that for RDs with masses of order $\sim0.30-0.35M_\odot$ and mass loss rates that characterize most of the nova accretion phase for most of the cycles, the effect of convection and bloating may be neglected, confirming the possibility of these systems entering long epochs of extremely low mass transfer rate --- hibernation.

\section*{Acknowledgements}
The support from the Authority for Research \& Development and the chairman of the Department of Physics in Ariel University are gratefully acknowledged.

\section*{Data Availability}
The data underlying this article will be shared on reasonable request to the author.

\bibliographystyle{aasjournal}
\bibliography{rfrncs}
\end{document}